\documentclass[12pt]{article}
\usepackage{amssymb}
\usepackage{comment}
\usepackage{graphicx}

\def\be{\begin{equation}}
\def\ee{\end{equation}}
\def\bear{\begin{eqnarray}}
\def\eear{\end{eqnarray}}

\newcommand\bra[1]{{\langle {#1}|}}
\newcommand\ket[1]{{|{#1}\rangle}}

 \def\IZ{\relax\ifmmode\mathchoice
 {\hbox{\cmss Z\kern-.4em Z}}{\hbox{\cmss Z\kern-.4em Z}}
 {\lower.9pt\hbox{\cmsss Z\kern-.4em Z}}
 {\lower1.2pt\hbox{\cmsss Z\kern-.4em Z}}\else{\cmss Z\kern-.4em Z}\fi}
 \def\IB{\relax{\rm I\kern-.18em B}}
 \def\IC{{\relax\hbox{$\inbar\kern-.3em{\rm C}$}}}
 \def\Ic{{\relax\hbox{$\inbar\kern-.22em{\rm c}$}}}
 \def\ID{\relax{\rm I\kern-.18em D}}
 \def\IE{\relax{\rm I\kern-.18em E}}
 \def\IF{\relax{\rm I\kern-.18em F}}
 \def\IG{\relax\hbox{$\inbar\kern-.3em{\rm G}$}}
 \def\IGa{\relax\hbox{${\rm I}\kern-.18em\Gamma$}}
 \def\IH{\relax{\rm I\kern-.18em H}}
 \def\II{\relax{\rm I\kern-.18em I}}
 \def\IK{\relax{\rm I\kern-.18em K}}
 \def\IP{\relax{\rm I\kern-.18em P}}

 \font\cmss=cmss10 \font\cmsss=cmss10 at 7pt
 \def\IR{\relax{\rm I\kern-.18em R}}

\def\dd{\mbox{d}}

\def\bra{\langle}
\def\ket{\rangle}

\def\D{\Delta}

\def\f{\phi}

\newcommand{\sm}[1]{\mbox{\scriptsize #1}}
\newcommand{\tn}[1]{\mbox{\tiny #1}}
\renewcommand{\@}[1]{\sqrt{#1}}
\renewcommand{\le}[1]{\label{#1}\end{eqnarray}}
\newcommand{\bea}{\begin{eqnarray}}
\newcommand{\eea}{\end{eqnarray}}

\newcommand{\eq}[1]{(\ref{#1})}

\def\ffract#1#2{\raise .35 em\hbox{$\scriptstyle#1$}\kern-.25em/
\kern-.2em\lower .22 em \hbox{$\scriptstyle#2$}}

\def\half{{1\over2}\,}
\newdimen\tableauside\tableauside=1.0ex
\newdimen\tableaurule\tableaurule=0.4pt
\newdimen\tableaustep
\def\phantomhrule#1{\hbox{\vbox to0pt{\hrule height\tableaurule width#1\vss}}}
\def\phantomvrule#1{\vbox{\hbox to0pt{\vrule width\tableaurule height#1\hss}}}
\def\sqr{\vbox{%
  \phantomhrule\tableaustep
  \hbox{\phantomvrule\tableaustep\kern\tableaustep\phantomvrule\tableaustep}%
  \hbox{\vbox{\phantomhrule\tableauside}\kern-\tableaurule}}}
\def\squares#1{\hbox{\count0=#1\noindent\loop\sqr
  \advance\count0 by-1 \ifnum\count0>0\repeat}}
\def\tableau#1{\vcenter{\offinterlineskip
  \tableaustep=\tableauside\advance\tableaustep by-\tableaurule
  \kern\normallineskip\hbox
    {\kern\normallineskip\vbox
      {\gettableau#1 0 }%
     \kern\normallineskip\kern\tableaurule}%
  \kern\normallineskip\kern\tableaurule}}
\def\gettableau#1 {\ifnum#1=0\let\next=\null\else
  \squares{#1}\let\next=\gettableau\fi\next}

\begin{document}

\begin{center}
\LARGE{Emergence in Holographic Scenarios for Gravity}\\
\vspace{1,2cm}
\normalsize{Dennis Dieks$^{1}$, Jeroen van Dongen$^{1,2}$ and Sebastian de Haro$^{2,3,4}$}\\
\vspace{0.8cm}
\small{$^{1}$\emph{Institute for History and Foundations of Science, Utrecht University, \\Utrecht, the Netherlands}\\
$^{2}$\emph{Institute for Theoretical Physics, University of Amsterdam, \\ Amsterdam, the Netherlands}\\
$^{3}$\emph{Amsterdam University College, University of Amsterdam,\\ Amsterdam, the Netherlands}\\
$^{4}$\emph{Department of History and Philosophy of Science\\University of Cambridge, Cambridge, UK}}
\end{center}
\vspace{0.4cm}

\begin{abstract}
\noindent
`Holographic' relations between theories have become an important theme in quantum gravity research. These relations entail that a theory without gravity is equivalent to a gravitational theory with an extra spatial dimension.
The idea of holography was first proposed in 1993 by Gerard 't Hooft on the basis of his studies of evaporating black holes. Soon afterwards the holographic `AdS/CFT' duality was introduced, which since has been intensively studied in the string theory community and beyond. Recently, Erik Verlinde has proposed that even Newton's law of gravitation can be related holographically to the `thermodynamics of information' on screens. We discuss these scenarios, with special attention to the status of the holographic relation in them and to the question of whether they make gravity and spacetime \emph{emergent}. We conclude that only Verlinde's scheme straightfowardly instantiates emergence. However, assuming a non-standard interpretation of AdS/CFT may create room for the emergence of spacetime and gravity there as well.
\end{abstract}

\section{Introduction}

During the last twenty years the concept of holography from quantum gravity research has grown into one of the key innovations in theoretical physics. By now it is studied in many diverse subfields and the literature on the subject has become enormous. One of the pioneering papers on holography, the article that
announced the celebrated `AdS/CFT' correspondence, has been cited more than ten thousand times.\footnote{(Maldacena, 1997).} Even fields that would seem far removed from quantum gravity are now engaging with holography. For example,
 central issues in condensed matter physics are addressed
using holographic ideas.\footnote{See for example (Hartnoll et al., 2008); (McGreevy, 2010); (Cubrovi{\'c} et al. 2009).} In short, the core idea of holography is that a lower dimensional quantum theory without gravitation (for instance, defined on the two-dimensional surface of a sphere) is capable of describing physical phenomena that include manifestations of gravity in a higher dimensional spacetime (such as the interior of the sphere).\footnote{For a
systematic statement of the holographic principle and appropriate choices of surface and interior, see (Bousso, 2002). For an early but comprehensive overview
of AdS/CFT, see (Aharony et al., 2000).}

It is time to pay attention to this important development also from the conceptual side: there are several
ideas here that relate not only to theoretical physics but also to more general foundational, conceptual and philosophical issues. Most importantly, holographic ideas clearly touch on philosophical questions of emergence and reduction.\footnote{See Rickles (2012) and Teh (2012). See also section 2.2.1 of Bouatta and Butterfield (2015), where additional reasons are provided why the time is ripe for philosophical assessment of these theories, despite the fact that they are not defined with the degree of precision that the mathematician would require.} Also in the physics literature these themes have come up, as reflected in some of the titles of articles on the subject:
these announce ``Emergent spacetime'', ``Emergent gauge fields'' or, e.g., promise a discussion of ``Aspects of emergent geometry in the
AdS/CFT context.''\footnote{(Seiberg, 2006), (Dom\`enech et al., 2010), (Berenstein and Cotta, 2006), respectively.} One of the publications
that we specifically focus on in this article is called ``On the origin of gravity and the laws of Newton."\footnote{(Verlinde, 2011).}

We will discuss a number of holographic scenarios and place them in the context of existing ideas about emergence.
It is not our aim to focus on a general analysis of the concept of emergence itself. Globally speaking, we sympathize with the characterization of
emergence as novel and robust behaviour relative to some appropriate comparison class,\footnote{(Butterfield 2011a, 2011b).} and we will use the term `emergence' accordingly.
What we wish to investigate here is whether, and if so how, recent holographic scenarios can be interpreted as representing such emergence, and whether one theory in a
holographic pair can justifiably be called more fundamental than the other. We will discuss three proposals in particular: 't Hooft's original formulation of the
holographic hypothesis, the AdS/CFT duality from string theory, and Erik Verlinde's recent ideas. Although these proposals are strongly interrelated, we will argue that
only Verlinde's account realizes emergence in a straightforward and uncontroversial way: gravity and spacetime here arise as thermodynamic phenomena in a coarse-grained
description. As far as we can see, the concept of emergence,
of higher dimensional gravity from lower dimensional non-gravitational processes, does not apply to AdS/CFT in its usual interpretation. However, we will argue that the analysis of Verlinde's scheme can cast new light on the interpretation of AdS/CFT, and we will accordingly suggest a way to create room for emergence also in that context.

That gravity perhaps originates from some deeper layer of reality and is different from other forces may intuitively be plausible to some extent, even if it is an intuition that has been alien to the string theory program and some of the other quantum gravity
programs.\footnote{Approaches that do assume that gravity originates from some underlying non-gravitational realm include those based on causal sets, group field theory, and tensor models. Our article will mostly focus on string and field theories                                                    .} Gravity distinguishes itself because it is universal:  it applies to all forms of matter and energy, and relates to the general framework of space and time itself---this may remind one of the universal
character of thermodynamic descriptions. Moreover, gravity is notoriously and essentially more difficult to quantize than other forces. This may suggest a difference of principle from the ordinary physical forces represented in the standard model. As already mentioned, studies of black hole physics have led to the hypothesis that quantum gravity theories within a volume correspond to theories \emph{without gravitation} on the boundary of this volume.
This seems only a small step from the notion that gravity \emph{emerges} from processes described by a theory without gravity;  it is this idea that we will critically analyze here.\footnote{These boundary spaces possess fixed spacetime geometries. These geometries could of course be considered as representing non-dynamical gravitational field configurations and therefore as manifestations of gravity in a restricted sense---but we will follow the tradition of calling them non-gravitational, in the way the Minkowski spacetime of SRT is usually viewed as not representing gravity.}

\section{The holographic hypothesis}

The central ideas of holography go back to the debates about the black hole information paradox that raged in the early 1990s. Important participants in these discussions were Gerard 't Hooft and Stephen Hawking; the latter famously claimed that black holes destroy information, which was opposed by the former.\footnote{(Hawking, 1976), ('t Hooft, 1985).} In 1993, almost twenty years after
the first results on the evaporation of black
holes had been announced by Hawking, 't Hooft put on the Los Alamos preprint server a short contribution to a future
Festschrift honoring the particle physicist Abdus Salam. It contained the first formulation of what would soon become known as the
\emph{holographic principle} of quantum gravity.\footnote{('t Hooft, 1993).}

In his article, 't Hooft made a programmatic start with the formulation of a unitary quantum theory of gravity, taking his cue from processes that he hypothesized to take place near black hole horizons. While leaving open what the exact degrees of freedom would be, 't Hooft argued via thermodynamical arguments that the entropy of a
black hole system is proportional to its horizon's area $A$.\footnote{A result that had earlier been argued for by Jakob Bekenstein (1973).} In natural units, and with the black hole's Schwarzschild radius given by
$2M$:
\begin{equation}
S=4\pi M^2=A/4. \label{bhent}
\end{equation}
This gives us a handle on \textit{how many} degrees of freedom there are in the black hole system, but it is also suggestive of the \emph{kind} of theory that should be able to describe these fundamental degrees of freedom. 't Hooft concluded that:
``The total number of [...] degrees of freedom, $n$, in a region of space-time surrounding a black hole is:"\footnote{('t Hooft,1993), p. 4.}
\begin{equation}
n = \frac{S}{\log{2}}=\frac{A}{4\log{2}}.
\end{equation}
Accordingly, there is a finite number of degrees of freedom in a black hole system.

't Hooft carried the argument one step further by pointing out that if a spherical volume $V$ is bounded by a surface $A$, the total number of possible states and the entropy inside $A$ are maximized if the volume
contains a black hole. Therefore, the number of degrees of freedom contained in any spatial volume is bounded by the size of its boundary surface area, and not by the size of the volume itself. In other words, there are many fewer degrees of freedom in the volume than one would expect on the basis of traditional calculations. So, ``we can represent all that happens
inside [the volume] by degrees of freedom on this surface [...]. This suggests that
quantum gravity should be described entirely by a topological quantum field theory,
in which all physical degrees of freedom can be projected onto the boundary. One Boolean variable
per Planckian surface element should suffice."

This statement contains the essence of the
\emph{holographic hypothesis}. Again 't Hooft: ``We suspect that there simply \emph{are} not more degrees of freedom
to talk about than the ones one can draw on a surface [...]. The situation can be compared with a hologram
of a three dimensional image on a two dimensional surface."\footnote{('t Hooft, 1993), p. 6.}

What does 't Hooft's account imply for the relation between the three-dimensional description and the surface description? The original 1993 text already suggests
some possible answers. 't Hooft's 1993 abstract states, interestingly, that at the Planck scale ``our world is not
3+1 dimensional." This appears to give precedence to the holographic description: the theory on the surface is more fundamental than the theory in the bulk.
However, 't Hooft's paper is not unambiguous on this point: in the same abstract,
he says that the observables in our world ``can best be described \emph{as if}"\footnote{('t Hooft, 1993), p.1,
our emphasis.}
they were Boolean variables on an abstract lattice (reminiscent of, e.g., a causal set approach), which suggests that the description on the surface only serves as one possible \emph{representation}. Nevertheless,
't Hooft's account more often assumes that the fundamental ontology is the one of the degrees of freedom that scale with the spacetime's boundary. In fact,
't Hooft argued that quantum gravity theories that are formulated in a four dimensional spacetime, and that one would normally expect to have a number of
degrees of freedom that scales with the volume, must be ``infinitely correlated" at the Planck scale. The argument is that the real number of degrees of freedom
is given by a theory on the surface, and because this number is much smaller that the number of independent degrees of freedom one could fit in the enclosed volume,
the volume degrees of freedom cannot be independent. 't Hooft even expressed the hope that this overdetermination might hold the key to an explanation of the notorious
EPR correlations.\footnote{See also e.g. ('t Hooft, 1999).}
The explanatory arrow here clearly goes from surface to bulk, with the plausible implication that the surface theory should be taken as more basic than the theory of
the enclosed volume. One is tempted to express this by saying that the space-time theory of the enclosed volume \emph{emerges} from the description on the surface. On
the other hand, the precise correspondence between boundary and bulk degrees of freedom does not immediately suggest the occurrence of new types of behaviour, which speaks
against emergence in the more specific sense of novel behaviour mentioned in section 1.

't Hooft proposed no concrete candidate for a theory on the surface. But given the above reading of his account, this surface theory---whatever it would be---would apparently be the best choice for a scientific realist who wishes to identify the fundamental objects in the quantum gravity world. So according to this reading there is no `ontological democracy'\footnote{A term proposed by E. Castellani at the 2012 Seven Pines Symposium; compare also Rickles (2011).} between surface and bulk.

Yet, there are also elements in 't Hooft's proposal that indicate a more equal status for
the bulk and boundary theories. Firstly, 't Hooft attributes the bulk theory a primary role when he points out that
it is its black holes that are responsible for the ``most direct and obvious \emph{physical} cut-off"\footnote{('t Hooft, 1993), p. 2,
emphasis as in original.} of the degrees of freedom, which explains the finiteness of the number of degrees of freedom.
Secondly, in the debate on
the information paradox, 't Hooft proposed that operators associated with observers moving inwards in
the black hole spacetime (in the bulk), and operators associated with observers that remain at a
distance, on the boundary, do not commute.\footnote{See ('t Hooft, 1996), pp. 30-34; 65-66.} This appears to point in the direction of a kind of complementarity between the two observers'
descriptions of the quantum black hole state.\footnote{The term `black hole complementarity' appears to have been introduced by (Susskind et al., 1993); for philosophical discussion, see (Belot et al., 1999), (van Dongen and de Haro, 2004).} `Complementarity' seems to imply that the two perspectives can claim equal rights in
describing the physics of the black hole. So, 't Hooft's holographic proposal wavers between boundary and bulk as fundamental ontologies. There is an interpretative tension here, that will resurface later in this article.

't Hooft's paper was programmatic and did not elaborate much on concrete possibilities for the bulk and boundary theories and their precise mutual relation. But the massive amount of later work on the so-called `AdS/CFT' duality has changed the situation. Here we have an example of a holographic relation between two theories that has been understood as a concrete instantiation of the ideas of 't Hooft (and others, in particular those of Leonard Susskind who followed up 't Hooft's work with an article that attracted considerable attention in the string theory community\footnote{(Susskind, 1995).}). We will discuss this concrete holographic proposal in the following
section. Let us end here by noting that soon after 't Hooft's paper, holography took on the role of a guiding principle in much quantum gravity work, not just in efforts based in string or field theory.\footnote{See e.g. its discussion in the book by Lee Smolin (2007, pp. 317-319), which is
quite critical of string theory, and advocates other approaches to quantum theories of gravity.}

\section{The AdS/CFT duality and its interpretation}\label{sectadscft}

We will first outline the AdS/CFT correspondence (3.1) and then discuss its interpretation, in particular with respect to issues of emergence and fundamentality (3.3). We also introduce the renormalization group (3.2), which is an important ingredient in AdS/CFT and also in Verlinde's scenario (to be discussed in section 4).

\subsection{What is AdS/CFT?}

The idea of a holographic correspondence between gravitational bulk theories and gravitationless theories defined on the boundaries of their spacetimes has found an explicit illustration in string theories in Anti-de Sitter (AdS) spacetime. There are reasons to believe that these string theories, which are meant to describe gravity, correspond exactly to Conformal Field Theories (CFT), without gravity, on the boundary of AdS.
This `AdS/CFT duality' was first conjectured by Juan Maldacena in 1997.\footnote{(Maldacena, 1997); an important early review article  of the subject is (Aharony et al., 2000).}

The AdS/CFT duality relates string theory in $d+1$-dimensional Anti-de Sitter spacetime (AdS$_{d+1}$) to a conformal field theory on a $d$-dimen\-sion\-al space
isomorphic to the boundary of AdS. The term `holography' was absent in Maldacena's original paper and initial excitement focused on the duality symmetry itself rather than its holographic aspects; the holographic nature of AdS/CFT was particularly highlighted in influential articles by Susskind and Edward Witten.\footnote{(Witten, 1998a; Susskind and Witten, 1998).}

Let us first look at some of the concepts that underlie the correspondence. Anti-de Sitter spacetime is the maximally symmetric solution of the Einstein equations with a negative
cosmological constant. In a suitably chosen local coordinate patch, its metric has the form:
\be
\dd s^2={\ell^2\over r^2}\left(\dd r^2-\dd t^2+\dd\vec{x}^2\right),
\ee
where $\vec{x}$ parametrizes $d-1$ spatial coordinates.
In these coordinates, distances diverge at the position $r=0$, which represents the boundary
of the space-time; this boundary is thus represented at a finite coordinate distance. The singularity in the metric at $r=0$ is therefore a large-distance singularity of the type one expects for spaces with
infinite volumes that are finitely parametrized.
The bulk metric induces a flat $d$-dimensional
Minkowski metric on the boundary at $r=0$, which is given by $-\dd t^2+\dd\vec{x}^2$, but only up to a conformal factor.
This metric is the metric of the fixed spacetime background of the conformal field theory.

A conformal field theory is a quantum field theory that is invariant under conformal transformations, that is, coordinate transformations that multiply the
metric by a scalar function (the `conformal factor' mentioned before). The fact that the bulk metric does not induce a Minkowski metric on the boundary of AdS,
but only a metric conformally equivalent to it, is therefore without consequences: because the boundary theory is a conformal field theory, it is insensitive to the conformal
factor of the metric.
In the standard example of AdS$_5$/CFT$_4$, in which $d=4$, the conformal field theory on the boundary is supersymmetric Yang-Mills theory
in 3+1 dimensions. The dual theory in this case is a type IIB string theory in AdS$_5\times S^5$ ($S^5$ is an additional internal manifold), which in the classical
limit reduces to supergravity in AdS$_5$.

The idea of a duality came to Maldacena when he was struck by the fact that there appeared to be two equivalent ways of describing, in the low-energy limit of small string length,
the states of a stack of $N$ D3-branes in the type IIB string theory (a `D3-brane' is a generalized type of particle solution, spatially extended in three dimensions; the `D' here stands  for `Dirichlet' since the D-brane is a surface on which a Dirichlet boundary condition is imposed on the string).
On the one hand, one can use the field theory living on the world volume mapped out by the D-branes. The branes are stacked close together, and their excitations
can be described by a Yang-Mills theory with gauge group SU($N$) at low values of the coupling constant, in which the number of branes determines the rank
$N$ of the gauge group. On the other hand, excitations in the  bulk geometry surrounding the D-branes can be described by using IIB supergravity in an AdS$_5\times S^5$
spacetime, evaluated in the regime of strong coupling. Maldacena took this correspondence between two equivalent descriptions as a hint that there existed a general and exact
relation between gauge and bulk theories at all values of the coupling.\footnote{(Maldacena, 1997).}

What this correspondence could look like was further investigated by, particularly, Steven Gubser, Igor Klebanov, Alexandre Polyakov and Witten.\footnote{(Gubser et al., 1998), (Witten, 1998a).} These authors proposed an exact equality between the partition function of the CFT (deformed by the insertion of an operator coupled to an external source) and the partition function $Z_{\sm{string}}$ of the quantum gravity theory in the AdS bulk. The partition function fully determines the expectation values of observables; so the claim that such an equality exists is a far-reaching hypothesis that is suggestive of some sort of physical equivalence. The precise form of the correspondence is given by:
\be\label{Zstring}
\Big\bra e^{\int\sm{d}^dx\,\f_{(0)}(x)\,{\cal O}(x)}\Big\ket_{\sm{CFT}}=Z_{\sm{string}}\left(r^{\D-d}\f(x,r)\Big|_{r=0}=\f_{(0)}(x)\right)~.\label{AdSCFT}
\ee
On the left hand side, ${\cal O}$ is an operator inserted via a space-dependent coupling parameter $\f_{(0)}(x)$. This coupling is not a quantum field, but can be thought of as representing a classical external source that probes the system. On the right hand side, the string partition function of the scalar field $\phi$ is computed with a prescribed boundary condition at $r=0$, given by $\f_{(0)}(x)$; $\D$ is a constant that depends on the dimension of the bulk spacetime and the field's mass.

The essential message of Eq.\ (\ref{AdSCFT}) is that there is a one-to-one correspondence between observables of the bulk theory (represented by fields) and observables of the CFT (operators). Given a boundary coupling parameter $\f_{(0)}$, associated with an operator $\cal O$ that couples to it, Eq.\ (\ref{AdSCFT}) enables us to calculate the bulk partition function for all bulk fields $\phi$ with $\f_{(0)}(x)$ as their boundary condition.\footnote{See for instance (de Haro et al., 2001).}

Full specification of the bulk theory determines, according to (\ref{AdSCFT}), the partition function of the CFT and therefore the expectation values of all observables of the CFT, since these can be computed from the partition function. Conversely, specification of the boundary CFT partition function leads to full knowledge of the partition function of the quantum gravity theory in the bulk. Although Eq.(\ref{AdSCFT}) only states the AdS/CFT correspondence for scalar operators, vector and tensor operators can be handled in a similar way. Equation (\ref{AdSCFT}) and its generalizations thus establish a one-to-one mapping between expectation values of observables of the two theories.
This is what we mean when we say that the AdS/CFT correspondence is a `duality'.

In discussions of duality, especially in the context of AdS/CFT, it is frequently stated that one is dealing with two theories that are the ``same.''\footnote{For example, (Aharony et al., 2000), p. 57, write: ``we are led to the conjecture that ${\cal N}$=4 SU($N$) super-Yang-Mills theory in 3+1 dimensions is the same as (or dual to) type IIB superstring theory on AdS$_5\times S^5$."} In his textbook on string theory, Barton Zwiebach described the situation thus: ``the term `duality' is generally used by physicists to refer to the relationship between two systems that have very different descriptions but identical physics."\footnote{(Zwiebach, 2004), p.\ 376.}
What such characterizations obviously aim at is the just-mentioned existence of a one-to-one correspondence between physical quantities (observables) and their expectation values, as well as between states, on the respective sides of the duality
(of course, `observable' is here used in its technical quantum mechanical sense and so refers to physical quantities that in principle could be measured; there is no direct relation to observability by the unaided human senses).

However, the classical actions of the theories are not the same, so that the full theoretical structures of the dual theories at least \emph{appear} different, which may explain the counter-intuitive element to the identity inherent to dualities.
For example, the line element of AdS, $\dd s^2={\ell^2\over r^2}\left(\dd r^2-\dd t^2+\dd\vec{x}^2\right)$, does not occur in CFT on Minkowski
spacetime, so that there is no manifest isomorphism between the mathematical structures of the theories in their standard formulations. Nevertheless, equation (\ref{Zstring}) ensures that
numerically correct accounts of any conceivable experiment or problem phrased with one theory's objects and concepts can be replicated using the concepts
and objects of the dual theory. It is remarkable that theories that  ``look very different,''\footnote{(Aharony et al., 2000), p. 60.} still yield the same numbers and that in this way a correspondence between amplitudes can be defined.
For this reason, some of the original protagonists of AdS/CFT found it comforting that the theories give these corresponding numbers in different ranges of expansion parameters: when calculations in one theory are made at strong coupling, the other theory should be considered at weak coupling, and vice versa.\footnote{(Aharony et al., 2000), p. 60: ``In this fashion we avoid any obvious contradiction due to the fact that the two theories look very different. This is the reason that this correspondence is called a `duality.' The two theories are conjectured to be exactly the same, but when one side is weakly coupled the other is strongly coupled and vice versa. This makes the correspondence both hard to prove and useful, as we can solve a strongly coupled gauge theory via classical supergravity."}

These considerations lead to the question of exactly \emph{how} different the dual AdS/CFT theories are; whether they share any structural properties apart from the one-to-one mapping between their observables and expectation values. Obviously, they should share all symmetries between observables. These correspondences between symmetries are indeed found in concrete examples. For instance, the space-time symmetry group of the CFT in $d$ dimensions (SO$(2, d)$)  equals the isometry group of $(d+1)$-dimensional AdS.
The theories also have a matching number of supersymmetries, and the internal manifold multiplying the AdS factor in the case of $d=4$,  $S^5$, corresponds to the
 SO(6) symmetry of the six scalar fields of $N=4$ super Yang-Mills. Moreover, both type IIB string theory and super Yang-Mills share a non-perturbative SL(2, $\mathbb{Z}$) symmetry.
These matching symmetries are generally taken as an indication that the AdS/CFT correspondence is \emph{exact}, and not only valid in a perturbative
approximation.\footnote{As suggested by e.g. (Green, 1999), (Bianchi, 2001), (Drukker et al., 2011).}

If the `field-operator' correspondence is indeed fully correct, then this suggests a physically meaningful mapping between the Hilbert space of string theory in the bulk and the Hilbert space of the CFT.\footnote{See (Aharony, 2000) pp. 90-98.} This would imply that a lot of physically significant structure is preserved when going from one theory to another, even in the absence of a full isomorphism between the two mathematical formalisms.
Nevertheless, there still is the possibility that the match between the theories may begin to fail at some order in the expansions and that as a result the duality may prove to be \emph{inexact}. This distinction between exact and inexact dualities is of importance for interpretational issues, as will become clear later on.

Regardless of whether an exact version of AdS/CFT holds true or not, it is clear that AdS/CFT relates bulk degrees of freedom, with gravity, to boundary degrees of freedom of a gravitation-less quantum field theory. So it is a concrete example (with specific characteristics) of 't Hooft's holography.

\subsection{The Renormalization Group and AdS/CFT}\label{renorm1}

The gravity side of the AdS/CFT duality suffers from large distance divergences; these correspond to high energy divergences on the CFT side. This is an example of a general feature: high energies on the CFT side (the ultraviolet or `UV' part of the spectrum) are related to large distances on the bulk side (the infrared or `IR' part), so that there is an `UV/IR correspondence'. Such divergences can be studied with the technique of the `renormalization group flow' (`RG flow'), which makes the effects of shifts in cutoff parameters explicit. This renormalization technique also plays an important role in Verlinde's scheme.

The  RG approach to renormalization, introduced by Ken Wilson in 1974,\footnote{See e.g. (Fisher, 1998) for the physics; (Hartmann, 2001), (Batterman, 2011) offer philosophical discussion.}  handles divergences and cutoffs differently from traditional renormalization procedures in quantum field theory. These traditional procedures typically introduced a cutoff in the integration range of a divergent integral, then performed a calculation (for instance of a path integral) and finally let the cutoff go to infinity. The novelty of Wilson's approach is the insight that there is no necessity to take cutoffs to infinity: interesting results can be obtained with finite values of the cutoff parameters. This approach disregards higher order quantum processes at energy scales that are above the cutoff value, but this is justified for processes that take place at low energies. Moreover, it is conceivable that completely new theories will be required to deal with processes at very high energies, and that these as yet unknown theories will solve the problem of the divergences: a situation which motivates leaving the cutoff at a finite value.

The renormalization group approach begins with limiting the integration range of momenta $k$ by introducing a cutoff $\Lambda$ in the partition function of the theory:\footnote{As $\Lambda$ is a momentum scale, not the integration range of the field, this step is indicated in a subscript added to the integration measure.}
\bea\label{pathi}
Z=\int[{\cal D}\f]_{0\leq|k|\leq\Lambda}\,e^{-S[\f]}.
\eea
Wilson's method essentially consists of repeatedly decreasing the momentum integration range by introducing a novel cutoff $b\Lambda$, while integrating out contributions to the path integral for $b\Lambda\leq|k|\leq\Lambda$. One repeats this process for smaller and smaller $b$, so $b\Lambda/\Lambda\rightarrow0$. To perform the path integral one splits the field $\f$ into Fourier modes $\f(k)$ with $0\leq|k|\leq b\Lambda$ and modes $\psi(k)$ with $b\Lambda\leq|k|\leq\Lambda$. The crucial point is now that the result of integrating out the modes $\psi$ can be represented by an adjustment of the parameters of, and the introduction of additional terms in, the original action. Writing the new (`effective') action as  $S_{\sm{eff}}$, we have:
\bea\label{partif}
Z=\int[{\cal D}\f]_{0\leq|k|\leq b\Lambda}[{\cal D}\psi]_{b\Lambda\leq|k|\leq\Lambda}\,e^{-S[\f;\psi]}=\int[{\cal D}\f]_{0\leq|k|\leq b\Lambda}\,e^{-S_{\sm{eff}}[\f]}.
\eea
Rescaling the momenta and coordinates, $k'=k/b$ and $x'=xb$, leads back to the original range $0\leq|k'|\leq\Lambda$ of \eq{pathi}---with a new, effective, action  that has `renormalized' couplings. The renormalized action may contain additional terms that were absent from the original form of the action. These additional terms represent the quantum effects of the high-energy modes that were integrated out in the renormalization step.

As mentioned at the beginning of this subsection, the high energy divergences on the CFT side correspond to large distance divergences on the gravity side of the duality. The just-described renormalization group procedure, which applies to the CFT divergences, accordingly has an AdS counterpart: we can introduce a cutoff at a large radial distance and take successive rescaling steps to smaller distances. In terms of the radial coordinate $r$ this means first introducing a small cutoff value $\epsilon$ (remember that $r=0$ corresponds to an infinitely great distance, i.e.\ the boundary of AdS; the introduction of the cutoff has the purpose of discounting the $r$ interval $(0,\epsilon)$ or, equivalently, distances greater than $1/\epsilon$.) This $r$ cutoff $\epsilon$ in AdS mirrors an UV (high energy) cutoff $\Lambda$ in the CFT. To implement the just-explained renormalization procedure on the AdS side we now introduce new cutoff values at greater $r$-values (and therefore smaller spatial radial distances) and integrate out the modes between the old and new cutoff values. We are thus moving inward from the boundary of AdS, in the sense of going to theories in which coarse graining has taken place over processes at greater distances.

Successive renormalization steps can be thought of as shifts in a space of theories. These shifts, and the corresponding integration over successive shells plus rescaling,  define the `renormalization group flow'.
If a theory is `renormalizable', the action does not acquire new terms under renormalization steps. An endpoint of a flow is a fixed point where the couplings no longer change so that the theory becomes scale-invariant.

These fixed points enable us to define \emph{universality classes} of theories.
Indeed, different theories may flow towards the same fixed point, which means that they show the same coarse-grained properties.
It may also occur that a given Lagrangian possesses several fixed points. It may have IR fixed points (i.e.\ fixed points at low energy-momenta), as just discussed, but it may also have UV fixed points, i.e.\ fixed points at high energies. In fact, semi-classical gravity near the boundary of AdS (i.e.~the IR) is related, by the UV/IR correspondence, to a UV fixed point of the corresponding conformal field theory. This makes the duality practically useful: calculations that are intractable in the UV boundary theory, may become tractable in the bulk where we just have semi-classical gravity.

Renormalization group transformations clearly involve statistical averaging: information is thrown away,
so that processes that are less relevant in the low energy regime are no longer described in a detailed way as the RG flow proceeds. This is a reduction of the fine-grained information available in the description. RG flow transformations in the space of theories can accordingly be conceived as steps toward
higher entropies.\footnote{(Gaite and O'Connor, 1995); (Swingle 2012).} This point of view will be relevant when we discuss Verlinde's ideas.

\subsection{Interpreting AdS/CFT}

Holography and duality raise interesting interpretational questions.\footnote{Useful discussions of duality include (Castellani, 2010); (Rickles, 2011).} Can one consider one of two dual theories as more fundamental than the other, so that it may become plausible to say that the description given by the less fundamental theory ``emerges''? Are we facing situations of empirical under-determination if there is no difference in fundamentality? After a preliminary look at possible reasons for favouring one theory over another in the context of AdS/CFT, we will attempt a more general appraisal of these questions.

\subsubsection{Is one side of the AdS/CFT duality more fundamental?}

One option is to consider the non-gravitational theory as more fundamental, and the higher-dimension\-al space-time and its gravitational
degrees of freedom as derived. We saw in section 2 that some of 't Hooft's intuitions went in this direction, when he introduced holography.
This viewpoint has the exciting consequence that spacetime (or at least some of its dimensions) would become non-fundamental: apparently (part of)
the spacetime description `comes from' a more fundamental description in non-spatiotemporal terms. For AdS/CFT this point of view has been
advocated by, for example, Seiberg (2007) and Horowitz (2005).

Seiberg (2007) has argued that spacetime cannot be probed at distances
smaller than a certain fundamental length scale, which according
 to him shows that spacetime cannot be part of a fundamental
description: at very high energies the notion of distance loses
 its meaning. However, apart from a general criticism one might
level against the step from verifiability to meaning, one should
note that in the case of an \emph{exact} duality and correspondence
between observables, a breakdown of empirical significance in the gravitational
theory (`GT') should be expected to be mirrored by a similar defect in the non-gravitational
theory (`NGT'), although at a different place in the theoretical structure. So it
is not clear at all how Seiberg's point about `loss of verifiability' implies a decision concerning fundamentality.

Horowitz (2005, p.\ 5) has proposed to consider the gravitational theory as {\it defined} through the NGT: ``since the
gauge theory is defined nonperturbatively [in AdS/CFT], one can view this as a nonperturbative and (mostly) background independent definition of string theory.'' Indeed, due to the UV/IR connection, the NGT is in fact the only available instrument to actually do calculations within the regime of strongly coupled quantum gravity. One should note, however, that
assigning precedence to the NGT based on this instrumental aspect has a pragmatic character, at least if one accepts that the duality between NGT and GT is exact. After all, an exact duality implies a one-to-one relation between the values of physical quantities, so that in this case it seems impossible to claim a \emph{descriptive} superiority of NGT over GT: as far as observables are concerned, the NGT and GT describe the physical world equally well or equally badly, even if one theory is more tractable than the other in a certain regime.

In the literature, one finds a near-unanimous consensus that the AdS/CFT duality should be taken to be exact, even
if there is not yet a proof of this exactness; many calculational results in concrete cases underwrite this consensus.\footnote{See for instance (Aharony et al. 2000).}
If the duality is \emph{not} exact, the question of the relative status of NGT and GT is relatively simple:
in this case the correspondence between observables can only be approximate, so that the straightforward question arises which one of the two theories is
better confirmed by experiment. The question of fundamentality in this case reduces to a question of empirical adequacy, even though an actual empirical verification
of the differences between the two theories might presently be out of reach.

If one of the AdS/CFT theories thus turned out to be more fundamental than the other on
 empirical grounds, it could of course be that the gravitational side is found to be the more fundamental one.
In this case, there would clearly be no reason to claim that spacetime and gravity emerge from the
boundary description.
For instance, it might  be
that exact duality fails in strong quantum gravity regimes, far from the semi-classical limit so that
strong quantum gravity phenomena cannot be captured by a CFT. The gravity side of the duality would in that case be superior,
in the uncontroversial sense of better fitting nature, even if it were convenient to employ the NGT as an instrument in
calculations. The NGT would then be a calculational tool of limited validity.

In spite of this possibility, in the literature one more often encounters the notion that spacetime and gravity are derivative and emergent
in some way.\footnote{This intuition has a long history.
Even Albert Einstein at some point expressed that if one desired a \emph{quantum} theory of gravity, one would have to
get rid of the spacetime continuum and thus arrive at a ``purely algebraic physics''; see his letter to Paul Langevin, 3 October 1935,
cited in (Stachel 1993), p. 285. Nevertheless, he preferred to stick with his own attempts at a continuum-based unification theory; see (van Dongen 2010), pp.\ 174-183.}
The difficulties that non-perturbative formulations of quantum gravity encounter, in combination with the universal character of gravity
that distinguishes it from other forces, may play a role in this expectation and in the relative  unpopularity of the point of view that
gravitation is fundamental. It should also be taken into account that statements about the emergence of space and gravity may often be meant in a straight-forward and metaphysically innocuous way: there is one spatial dimension \emph{more} in the bulk theory, which has thus been ``created'' or has ``emerged'', without a clear commitment to the idea that the CFT side is metaphysically more fundamental.\footnote{We have not found, in the AdS/CFT literature, any explicit statements that gravity should be considered as
more fundamental than gauge theories. However, there are numerous articles in which the gauge theory side of the duality is used as a {\it tool}
for predicting bulk physics, while the latter seems implicitly assumed to be more fundamental: boundary calculations are here treated merely
instrumentally. See for instance the discussion of the
Big Bang scenario by Hertog and Horowitz (2005).}

\subsubsection{Duality, fundamentality and emergence}\label{internalist}

As is clear from the above introductory remarks, the distinction between exact and approximate dualities is important for the question of differences in fundamentality of the two sides of a duality. In the case of approximate duality there is no complete empirical equivalence between the theories in a dual pair, so that uncontroversial criteria for theory evaluation can be deployed. In this situation there is scope for the notion of `emergence': the duality now boils down to an inter-theoretic relation that could resemble the one between thermodynamics and statistical mechanics, in which one description approximates the other. In this analogy it is undisputed that the atomistic description is the more fundamental one, even if in most practical situations it is impossibly inconvenient to take recourse to calculations on the micro-level---this latter circumstance is relevant to the pragmatics of the situation, not to considerations about fundamentality. Continuing the analogy: on the thermodynamic level concepts like `temperature' and `pressure' become applicable---concepts that capture objective aspects of physical reality, even though they cannot be applied to the more fundamental atomistic description. `Temperature' and `pressure' can here be said to emerge in a clear and uncontroversial sense of emergence: these concepts figure in the characterization of novel and robust behaviour that is insensitive to the underlying atomistic and molecular details. This emergence involves an asymmetry between the theories that are involved: thermodynamics emerges from the micro description, but not the other way around. A relation of approximate duality might well be similar in relevant aspects and give rise to an effective description that emerges from the more fundamental theory in the dual pair; as we will see in section 4, the situation in Verlinde's proposal can be considered as a case in point.

However, most discussions about duality and its philosophical consequences take place against the background of the assumption that the duality is exact, and here it is less clear how we should judge the relative status of the theories that are involved. By definition there is in this case a precise one-to-one mapping between the observables and their values in the two theories. This suggests that the theories are empirically equivalent: for each physically significant number in one theory there is an exact counterpart in the other.

A natural objection is that a one-to-one mapping between physical quantities and their values by itself does not imply empirical equivalence, on the grounds that the mapping may relate different quantities, with different physical meanings, and different regimes of coupling strengths. This objection presupposes that the physical meaning of the quantities in each of the theories has been fixed independently; that we already know what terms such as `energy' and `distance' mean (in the sense of the reference of these terms in physical reality) in both theories before we start contemplating the relation between the theories. This will be the case if there exists what one might call an `external point of view', from `outside' the two theories, from which the reference relation between each of the theories and physical reality can be defined. If this is the situation that is being considered, duality between theories expresses a \emph{symmetry} in the physical world: exactly the same relations that obtain between, say, the energies in certain processes also obtain between, e.g., distances in certain other processes.

An example of this kind is provided by the source-free Maxwell equations, which exhibit perfect symmetry between the $E$ and $B$ fields. When we consider the application of these equations to a source-free region of space, the form invariance under an exchange of $E$ and $B$ reflects a physical symmetry that is present in this region, due precisely to the absence of charges. If other regions do contain charges, this breaks the symmetry and determines unambiguously which physical fields the $E$ and $B$, respectively, refer to. In this case exchanging $E$ and $B$ in the source-free region does not change anything in the local form of the equations, but it does imply a drastic change in the physical situation that is described: electric fields are replaced by magnetic fields. So here the duality connects \emph{different} aspects of the world (the electric and magnetic fields in a charge-free region) that possess an isomorphic internal structure. Situations of this kind enable us to make \emph{models} for physical phenomena falling under one theory with the help of concepts from another theory (as in the case of hydrodynamic models for electrostatics).

Clearly, in this situation the notion that the duality is connected with emergence does not even suggest itself. With regard to the example: the symmetry between $E$ and $B$ does not entail anything about a possible origin of electricity in magnetism or the other way around. Similarly, reflection symmetry in ordinary space has no implication for a possible emergence of `left' from `right' or \emph{vice versa}.

The situation becomes more interesting, and more in the spirit of discussions about duality in the context of present-day fundamental physics, in the case of a global and exact duality between two theories that are both candidate descriptions of the same world (including experiments and their outcomes)---as in the example of holographic pairs. In this situation, it is no longer clear that there exists an `external' point of view that independently fixes the meanings of terms in the two theories. Think of AdS/CFT: although we speak about `energy' in CFT, the very idea of holography is to represent, by means of this CFT, the distance relations in the bulk theory, which involves a strict correspondence between CFT-\emph{energies} and bulk \emph{distances}. In other words, if the holographic idea is to work we should assume that energy in CFT and distance in the bulk refer to the same thing.

Generally speaking, in duality cases like this we are dealing  with two structures of observables and their (expectation) values that have exactly the same internal relations to each other in the two respective theories. Without an independent external viewpoint, the only thing to go on with regard to the meaning of these observables is now how they are positioned within their two respective networks of relations. But this means that we are justified in concluding that the isomorphism between the structures of observables can be cashed out in terms of \emph{equality} rather than \emph{symmetry}. The symbols used in the two theories may be different, but in view of the identical roles quantities play in relation to other quantities, and the values they assume, \emph{identifications} can be established: $A$ in one theory will denote exactly the same physical quantity as $B$ denotes in the other if these quantities occupy structurally identical nodes in their respective webs of observables and assume the same (expectation) values.\footnote{What is assumed here is not a structuralist doctrine about what the world is like, but rather a view about how a mathematically formulated `theory of everything' can correspond to the world if no \emph{a priori} `rules of correspondence' between the theory and world are given. In this case it is only the internal structure of the theory that can decide how it can be applied to the world. So the structuralism here is epistemic. Even when correspondence rules for one of the two theories are given, as in the previous paragraph, the duality map induces a {\it second} set of correspondence rules between the symbols in the second theory and the {\it same} physical quantities.}

We are thus dealing with a very strong case of empirical equivalence: the substructures of observables of the two theories coincide. In the recent philosophical literature about empirical equivalence and under-determination (for the greater part responding to the seminal paper by Laudan and Leplin) the possibility of such a thorough-going empirical equivalence is often doubted.\footnote{See (Laudan and Leplin, 1991); for a recent critical discussion see (Acu\~{n}a and Dieks, 2014).} But the case of exact duality resists most of the usual arguments. For instance, in examples of (potential) duality discussed in the literature it certainly is not true that one of the two theories fails to meet standards of theoreticity, or is an artificial parasite on the other. Such standard objections against empirical equivalence have the purpose of removing the threat of theoretical under-determination, the dilemma that arises when it is impossible to reach an empirically justified choice between theories. Exact duality appears to revive this threat, by avoiding the standard objections against putative cases of empirical equivalence.

In the philosophical literature, it is usually argued that it would betray a superficial instrumentalism to \emph{identify} two empirically equivalent theories---to consider them as just variant formulations of one and the same theory. The thought behind this objection is that differences in theoretical structure between theories may well correspond to differences in physical reality, even if these differences are not (yet) observable: we should not assume that the descriptive physical content of theories is exhausted by the theory's observable consequences. But in our cases of exact duality the situation is different from what is usually assumed in these philosophical discussions of empirical equivalence. As we have pointed out before, the `observables' that are in one-to-one correspondence with each other in cases of exact duality are not defined via a notion of observability as in the debate about empiricism and scientific realism. Rather, they stand for what is physically real and meaningful according to the theories under discussion (i.e.\ expectation values of all physical quantities), even if there are no possibilities of direct observation. So what we are facing is not the standard situation of empirical equivalence in which two different physical theories coincide `on the surface of observable phenomena': we are dealing with theories that coincide exactly on everything they deem physically real.\footnote{Compare with Matsubara (2013).}

We therefore conclude that in the case of an exact duality between theories without fixed external rules of correspondence a very strong form of equivalence arises; but one that does \emph{not} lead to theoretical under-determination. Because it is inherent in the notion of exact duality in this case that the two theories completely agree on everything that is physically meaningful, the two sides of the duality should be taken as different representations of one and the same physical theory. The two theories collapse into one; and there is consequently no emergence of one side of the duality from the other.

In summary, emergence is a potentially applicable notion when we are dealing with approximate duality. In this case one theory may be uncontroversially more fundamental than the other, and the relation between the two may be similar to the one between thermodynamics and statistical mechanics. But in the case of an exact duality the situation is different. If it is independently given what the physical quantities in the two theories stand for (an `external' viewpoint) then it is to be expected that the two theories identify different parts of the  physical world that possess the same internal structure. In this case the duality expresses a physical symmetry; and there is no implication of emergence. If no external viewpoint is available for at least one of the theories, so that the physical meaning of theoretical quantities has to be determined from the roles they play within their theoretical framework, the natural conclusion to draw from an exact duality is that we are dealing with two formulations of the same theory. This is the situation we will encounter in the case of holography---there is no emergence here of one theory from a holographic dual.

\subsubsection{Renormalization and emergence in AdS/CFT}

When we are dealing with \emph{one} theory in different formulations, there could still be emergence \emph{within} this theory. For example, there could be an effective macroscopic description if the system possesses very many degrees of freedom.
This could justify speaking about emergent macro-behaviour. Could Einstein gravity be emergent in this way in AdS/CFT?  In the vast majority of actual examples of AdS/CFT one relates a {\it semi-classical} bulk theory\footnote{This is the approximation in which the string length is small compared to the AdS radius.}
 to a
CFT that is considered at large $N$, i.e.\ for high values of the rank of the gauge group SU$(N)$. The latter represents a particular kind
of (semi-)classical limit.\footnote{Technically, for the case of a four-dimensional CFT, one takes $N$ to be large but keeps the
product $g^2N$ fixed, where $g$ is the coupling constant.
The quantity that is held fixed is called the `'t Hooft coupling.'
In this limit, only a limited class of Feynman diagrams (called `planar' because they
can be written on the plane) contribute to the observables, and these diagrams are generally
reproduced by the saddle point of a classical theory. For a philosophical introduction to this aspect, also focused on the topic of emergence, see Bouatta and Butterfield (2015).} This gives rise to the question of whether gravity as we know it from
classical theories is fundamental. It has indeed been suggested (in fact, in many quantum gravity approaches) that the metric field (the central quantity
in general relativity) will not be one of the fundamental fields in a fundamental microscopic bulk theory but will somehow appear in a limit arising from a fundamental
microscopic theory that has very different space-time properties.
In the context of string theory, where AdS/CFT has been proposed, a concrete result that supports this idea is that the metric is reproduced in the regime of very small string length.\footnote{See e.g.
(Green et al., 1987), p. 115, where an excitation of the string is found that corresponds to a nearly flat ambient metric. This also requires small string coupling.
Increasing the string coupling allows for more highly curved metrics, see ibid.\ pp.\ 166-183.}

One might thus speculate that ordinary space-time concepts in the bulk only make sense after taking some limit; and there are indications that this limit may be, in some generalized sense, thermodynamic. Indeed, progressively neglecting quantum corrections to the Einstein equations in the bulk corresponds,
via the AdS/CFT duality, to renormalization transformations in the CFT (see sect.\ \ref{renorm1}) that throw out higher order terms in the
action.\footnote{See (de Boer et al., 1999).} If we interpret this sequence of coarse-graining/renormalization steps as the
transition to a thermodynamic limit, we see how a thermodynamic limit on the boundary may be associated with the emergence
of classical Einsteinian gravity in the bulk. In such a scenario, gravity is \emph{not} emergent \emph{due to duality} but rather because of coarse-graining and the existence of a
huge number of degrees of freedom. We will return to this point of view later on, after we have introduced Verlinde's ideas.

\section{Gravity as an entropic force}

The third holographic scenario that we want to analyze in some detail is the recent explanation for gravity proposed by Erik Verlinde. We first list its key assumptions, and add details in section 4.2. We will try to disentangle the logical structure of Verlinde's argument and assess some of its conceptual and interpretative consequences in section 4.3.

\subsection{Holography and Newton's law of gravitation}

Verlinde invites us to consider a closed two-dimensional space, e.g.\ the surface of a sphere, on which a quantum theory is defined. He remains quite unspecific about this quantum theory (the possibility of doing so is one of the salient points in his proposal). It is sufficient to assume that the theory describes physical processes on the surface that can be characterized in a very general information-theoretic way, as `changes in information'. More concretely, the surface area of the sphere is imagined to be divided in small cells, each of which can contain one bit of information. A physical state corresponds to a distribution of 0-s and 1-s over these cells, and time evolution of this state corresponds to a change of the distribution.
The holographic principle now suggests that such a surface theory can also represent physical processes that go on \emph{inside} the sphere. In particular, the bits on the surface may encode where matter is located in the interior.

The number of bits on the sphere is taken to be very large, which makes it possible to assume that an effective thermodynamic description can be used instead of the original quantum theory defined on the micro-level of cells and bits. From the viewpoint of thermodynamics the physical processes that take place on the surface (changes in the 0-s and 1-s in the cells) can be characterized as processes that maximize entropy: the distribution of 0-s and 1-s tends to an equilibrium distribution.

The core of Verlinde's proposal is that this entropic process on the surface corresponds, via the holographic principle, to \emph{gravitational} processes in the bulk (the interior of the sphere). In other words, the changes in 0-s and 1-s on the surface, described in the thermodynamic regime, yield a description of matter in the interior that is falling inward. This idea is made plausible through a simple deduction of Newton's law from a holographic translation of thermodynamical equations on the surface---we will reproduce this derivation here.

First, it is assumed that there is a number of active bits, $N$, which is proportional to the sphere's total surface area $A$ (constants are introduced and handpicked for later
convenience\footnote{With this choice of constants, the number of bits is the area of the surface measured in Planck units.}):
\bea
N = \frac{A c^3}{G\hbar} \label{holo}~,
\eea
in which the area of the sphere can be written as
\bea
A=4 \pi R^2 .\label{surface}
\eea
Although the terminology that is used suggests otherwise, no embedding in a three-dimensional world is presupposed: in the surface theory $R$ should be considered as a quantity that is \emph{defined} by Eqs.\ (\ref{surface}) and (\ref{holo}).
As already announced, it will be a central assumption in the derivation of Newton's law that the thermodynamic limit can be taken on the surface. A \emph{temperature} $T$ will therefore be definable, and it is assumed that in thermodynamic equilibrium this temperature relates to the energy $E$ through the law of equipartition:
\bea
E=\half Nk_{\tn B}T~.
\eea
We can now \emph{define} the quantity $M$ by:
\bea
E=Mc^2.
\eea
On the surface, $M$ is just an alternative expression for the thermodynamic energy; but via the holographic correspondence it will soon acquire the interpretation of the total gravitational mass that is present in the interior.

As we have seen in the discussion of the Renormalization Group (see section \ref{renorm1}), the quantum theory on the surface can be subjected to coarse-graining renormalization steps. Think, to make this more concrete, of the renormalization of a theory that describes bits as quantum spins on a lattice. The renormalization steps take lattice cells together and average over them; in this way they produce an increase in lattice-cell sizes,\footnote{On these `block spin transformations', see e.g. (Fisher 1998), pp. 666-669.} and thus effectively reduce the area of the surface (after rescaling the size of the cells). Going to a more coarse-grained description, and therefore discarding part of the fine-grained information (in the high energy modes of the field theory), is thus equivalent to considering a quantum theory on a smaller surface (the `screen') than we had before (see Figure 1; we will return to this). Using the relation (\ref{surface}) between surface and $R$, now interpreted as radial distance, we are led to a representation in three dimensions by means of a nested set of spheres, all with the same center but with different radii. In this way it becomes possible to talk about both the inside and the outside of any given sphere, although all quantities were defined within two-dimensional theories.

\begin{figure}[top]
\begin{center}
\includegraphics[height=7cm]{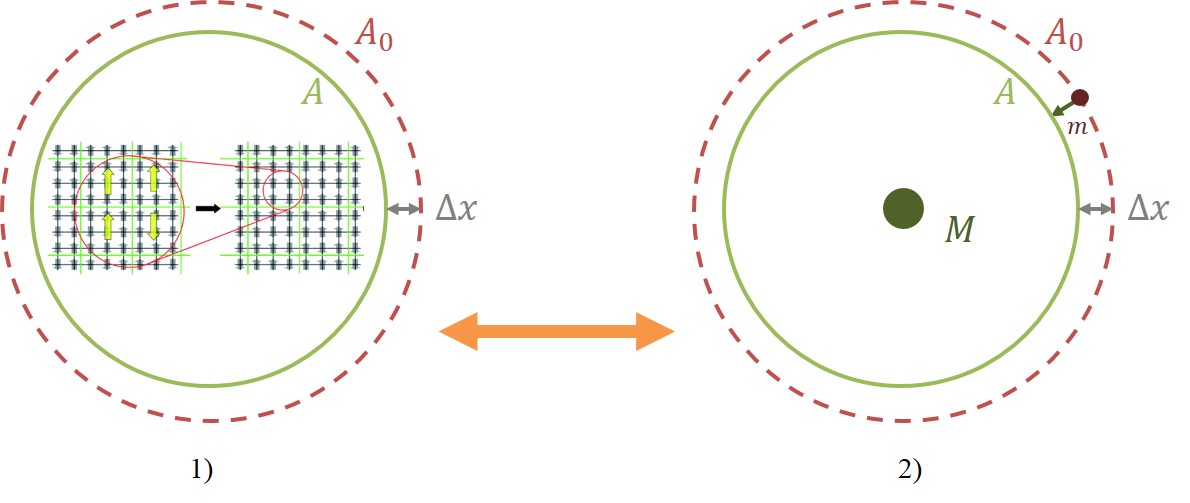}
\caption{\small{A test mass in three-dimensional space, close to the position of one of the spheres (`screens'), feels a force (on the right). This force, according to the holographic hypothesis, can be seen as an expression of the tendency towards increasing entropy on the screen. The growth in entropy in time allows for increasingly coarse-grained descriptions realized by renormalization transformations in the surface system, as depicted on the left;  renormalization group transformations are implied when the screens `follow the particle inwards'. On the sphere on the left, an example is sketched of `integrating out' degrees of freedom via an RG-like step, i.e., a block spin transformation: groups of microspins are identified with an overall spin value, after which the lattice is reduced in size, along with the reduction in surface area of the sphere. Such a transformation is associated with an increase in entropy, as `micro-information is thrown out', i.e.\ in the coarse-grained description single spin values in fact represent more spins on a `finer-grained' level. This microscopic structure becomes irrelevant when the system relaxes.}}
\label{Verlinde-figure}
\end{center}
\end{figure}

Now imagine that in this external bulk description (in three-dimensional space) a particle with mass $m$ is represented as coming from outside a screen and as changing its distance with respect to it by $\Delta x$. This change in distance is measured in time; and the time parameter is shared between the bulk and the surface descriptions. Via holography, the bits on the screen encode everything that is going on within the screen's interior. After the approach of the particle, and its subsequent fall into the interior, there is more matter inside and therefore more has to be encoded on the surface, so the number of active bits on the screen increases. The increase of the total mass must also correspond to an increase in the entropy. In analogy to, and generalizing upon Bekenstein's ideas about entropy changes when masses fall into a black hole---but without assuming a priori anything about a gravitational origin---Verlinde takes this change in entropy to be:\footnote{The numerical factor relating $\Delta S$ and $\Delta x$ is fixed by considering a thought experiment first worked out by Bekenstein in the 1970s in the context of black holes: Bekenstein had argued that when a particle is added to a black hole, the latter's area increases by, minimally, $8\pi\hbar$, which can be added when the particle is at one Compton wavelength from the horizon (Bekenstein 1973, p.\ 2338). Arguably, it can then no longer be distinguished from the black hole. In the same way, if the particle is at the distance of its Compton wavelength from the screen, the entropy on the screen is raised by one bit (with a factor of $2\pi$ put in by hand): $\Delta S=2\pi k_{\tn{B}} \, \mbox{when} \,  \Delta x={\hbar\over mc}$.
Generalizing for arbitrary distances leads to relation (\ref{entropy}), with ${mc\over\hbar}\,\D x$ being the distance expressed in units of the Compton wavelength.}
\bea
\Delta S = 2 \pi k_{\tn B}\, \frac{mc}{\hbar}\, \Delta x~. \label{entropy}
\eea
In the thermodynamic description on the screen the process of falling inward is described as an approach towards equilibrium, in which the entropy grows. Such processes can be characterized phenomenologically as the result of the operation of an effective `entropic force' $F$ that represents the effects of changes in entropy:
\bea
F\,\Delta x=T\Delta S~. \label{force}
\eea

The peculiarity of an entropic force is that it does not derive from an interaction, but arises from the statistics over microscopic degrees of freedom resulting in thermal motion. A typical example is the force that can be used to describe the behavior of a polymer, stretched in the direction $\Delta x$. On the fundamental level, viz.\ the level of the atoms that make up the polymer, there need not be any force: the polymer may consist of short chains of atoms that are connected but can rotate freely with respect to each other. However, as a result of random microscopic motion, the polymer will with overwhelming probability end up in a macroscopic state that corresponds to a large phase space volume; this will be a state in which the polymer is coiled up (there are vastly more coiled-up microstates than states in which the chains of the polymer are collinear). So from the macroscopic point of view a definite directedness in the behaviour of the polymer manifests itself: it tends to coil up, in spite of the microscopic randomness. This tendency (associated with an increase in entropy) can be phenomenologically described as caused by an elastic force obeying Eq.\ (\ref{force}).

Going back to our case in which the growth of entropy is associated with particle motion in the bulk, we can determine the magnitude of the effective force. Since the time parameter is shared between bulk and surface descriptions, and the bits under consideration correspond (see \eq{entropy}) to the mass $m$ whose change in position $\Delta x$ we are considering, the force must apply to the mass $m$ in the bulk. 
Simply combining the above relations (\ref{holo})--(\ref{force}) yields the result
\bea
F=G\, \frac{Mm}{R^2}~. \label{Newton}
\eea
This suggests that gravity is an \emph{entropic force} whose ``corresponding potential has no microscopic meaning''.\footnote{(Verlinde, 2011) p.\ 4.
The force is here related to the acceleration in the usual way; acceleration itself will be related to an entropy gradient (see also our next section).} In his paper Verlinde shows that it is possible to give a similar derivation of the Einstein equations.\footnote{A related derivation of the Einstein equations was given earlier by T. Jacobson (1995), who also already claimed that they are ``born in the thermodynamic limit" (p. 1260). Various approaches to quantum gravity have also proposed and elaborated that gravity should be studied as an emergent phenomenon in this sense; see e.g.\ (Barcelo et al., 2001), (Konopka et al., 2008), (Hu, 2009), (Oriti, 2014),  (Padmarabhan, 2015). Verlinde says that his innovation particularly lies in the explicit interpretation and construction of gravity as an entropic force (2011, p. 9), and the discussion of this in the context of string theory and AdS/CFT. \label{alternatives}}

So new space dimensions and gravity may correspond to things happening in a lower-dimensional and non-gravitational background: in the above account the third dimension appeared as a coarse-graining parameter of the surface theories, and gravitation came in as the three-dimensional translation of the coarse-graining and the associated thermodynamic description of what happens on the surface. Indeed, Verlinde states that he has ``reversed the arguments'' that have yielded holography and black hole thermodynamics, so that from holography and thermodynamics we now can `understand' gravity: this has ``shed new light on
the origin of gravity.''\footnote{(Verlinde, 2011) p.\ 9.}
But there are questions about the status of the various assumptions that have been made. And, most important for our purposes: is it justified to say that in this scenario the surface theory is \emph{more fundamental} than the bulk theory, so that the surface theory may be called the \emph{origin} of gravity?

Two ingredients in Verlinde's proposal are essential for the derivability of gravity from the gravitation-less surface theory, namely 1) the holographic correspondence between surface and bulk, and 2) the transition from the microscopic to the thermodynamic mode of description, which grounds the characterization of gravity as an entropic phenomenon. Verlinde speaks about the emergence of gravity and space without differentiating these two relations too strictly; however, understanding the difference between them is important. We will therefore discuss these two core assumptions in more detail, first following a standard line of thought (in section 4.2) and then offering a new interpretation (section 4.3). This will also cast new light on emergence in AdS/CFT.

\subsection{The correspondence between information loss and gravity}\label{4.1}

\begin{figure}[top]
\begin{center}
\includegraphics[height=6cm]{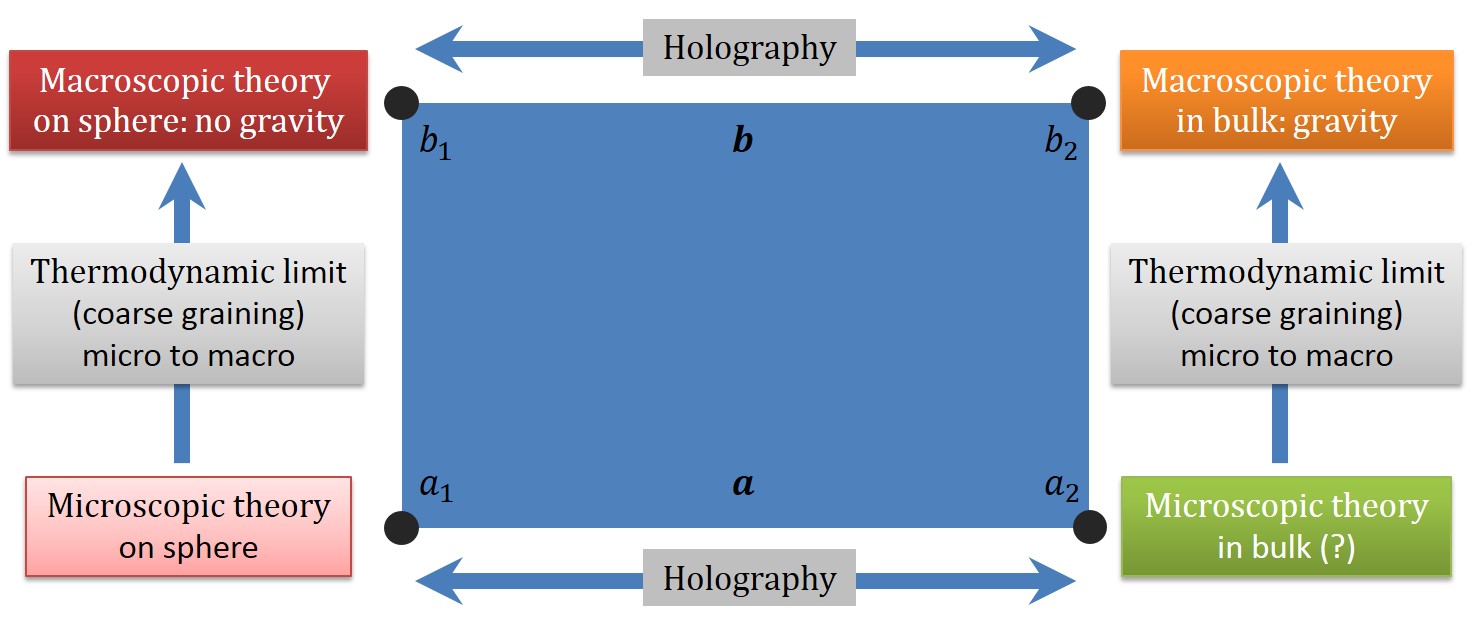}
\caption{First interpretation of Verlinde's scenario for holography and emergence: at the microscopic scale, one assumes that a holographic and exact duality {\bf a} holds between a screen quantum theory \emph{a1} and a bulk quantum theory \emph{a2}. For large numbers of degrees of freedom, coarse-grained descriptions suffice for capturing the relevant physics; the microscopic duality plus the thermodynamic limit give a new holographic relation {\bf b} between a thermodynamic description of the system on the screen, \emph{b1}, and gravity in the bulk, \emph{b2}.}
\label{Micro-macro}
\end{center}
\end{figure}

There are two holographic correspondences to consider, one at the micro-level, \textbf{a}, and one at the macro-level, \textbf{b}.
The first correspondence we consider, {\bf a}, is the holographic identification of degrees of freedom between two microscopic systems \emph{a1} and \emph{a2} (see Figure \ref{Micro-macro}).
The system \emph{a1} is defined on the surface and is described by a theory without gravity. As we have seen, the microscopic dynamical details of this system are not relevant in Verlinde's proposal; we only specify that it should be possible to speak about the thermodynamic regime, the number of degrees of freedom of the theory and the statistical interpretation of entropy---this is sufficient for the argument to take hold. The system \emph{a2} is a system of masses in the bulk. In contrast to what is often assumed in discussions of the holographic principle, in this scheme we do not need to assume that \emph{a2} is described by a microscopic theory including gravity in any traditional sense (either a quantum gravity theory, or Einsteinian gravity). The bulk microscopic theory dealing with \emph{a2} could be without gravity in a recognizable form because in Verlinde's scheme gravity as we know it is taken to arise from thermodynamics.\footnote{One may ask when exactly microscopic interactions count as `gravity' (for instance, a microscopic force mediated by a spin two excitation connected with diffeomorphism invariance might already qualify as gravity). For the sake of our analysis, we will use the term `gravity' in a restricted sense, namely either as Newtonian or Einsteinian gravity.}

Verlinde does not state explicitly which precise form of holographic relation between \emph{a1} and \emph{a2} he has in mind. However, his various remarks and his use of string duality parlance indicate that, initially at least, he is thinking of a \emph{bijective map} between a microscopic quantum theory in the bulk and a microscopic quantum theory on the screen, so that we have an exact duality.  We will return to this later.

The correspondence {\bf a} leads naturally to the consideration of a second correspondence relation, {\bf b}, between two systems \emph{b1} and \emph{b2}. These two systems are the
macroscopic thermodynamic versions of the systems mentioned under {\bf a}. The system \emph{b1} is the system described by the surface theory again, but now considered in the
regime in which the number of degrees of freedom is very large and an effective macroscopic description can be employed (see Figure 2). This transition to a macroscopic
description consists in the `throwing away' of irrelevant degrees of freedom, which can formally be represented by RG transformations, block spin transformations (see section 4.1), or similar coarse-graining steps. In the surface language: when fewer degrees of freedom are explicitly taken into account, a smaller surface suffices for the description (a surface with fewer cells, and therefore
with less information carrying bits).
As macrostates that can be realized by more microstates are more probable, the system will move towards them, as in the analogy of the released polymer. While the system evolves in time
to states that are closer to equilibrium, it becomes equally well describable, from the macroscopic point of view, by fewer microscopic variables; less `information' is relevant to
the macroscopic description of the system, and in this way entropy grows---not only in the bookkeeping of coarse-graining, but also in time.

As we saw before, a concrete example of how this works is given by a screen theory that describes a large number of quantum spins on a lattice. While the spin system diffuses naturally, it can be characterized by increasingly coarse-grained theories: successive block spin transformations lead to more efficient theories, mentioning fewer degrees of freedom but equally well suited to describe the system.
As explained in section 4.1, the renormalization steps produce a smaller copy of the screen with less microscopic information. In a three-dimensional picture the new screen can be imagined as placed inside the original one; the striving for thermodynamic equilibrium on the screen then corresponds to the adequacy of using a succession of increasingly coarse-grained theories, defined on smaller and smaller spheres with shrinking interiors.\footnote{Note that the \emph{maximum} possible amount of entropy
\emph{decreases} as the surface decreases. In the gravitational correspondence that we are discussing, the screen capacity cannot be further reduced when we reach the horizon of a black hole. This is the final equilibrium situation in which a further growth of entropy is impossible. Until that point has been reached, however, `reducing screen size by a renormalization group step' or considering successively smaller screens can go hand in hand with increasing physical entropy.}

The system \emph{b2} is the system described by the holographic counterpart of the theory describing \emph{a2}, again in the thermodynamic limit. As we have seen, Verlinde's central claim is that this macroscopic bulk theory describes the interior of the spheres in terms of masses and gravitational forces between them: it is a gravitational theory. Via the holographic correspondence \textbf{b} (conceived as a bijective mapping) \emph{b1} and \emph{b2} become two alternative ways of describing the same thermodynamic system. For example, the parameter $x$ (Eq.\ \ref{force}) is defined as a cut-off parameter in the surface theory, which keeps track of coarse-graining steps on the surface, but it becomes an added spatial dimension in the bulk description. However, in both cases it figures in the same formulas so that structurally the descriptions are the same. The just-described process of entropy growth on the screens is thus represented as a gravitational process that needs less and less space for its description---because masses fall inwards during their approach to gravitational equilibrium---and in which the gravitational bulk properties correspond one-to-one to thermodynamic quantities defined on the screens. Because the change in the $x$ of the mass in the bulk is related to the change in entropy on the surface, an appropriate evolution of the entropy with respect to the common time, will correspond to an acceleration of the mass proportional to the force. So, for example, the gradient of the gravitational potential in the bulk turns out to track the level of coarse-graining of the surface theories; the force felt by a test particle in the bulk in this way encodes the entropy gains on the surfaces.

One argument in favour of these ideas is that usually in the thermodynamic regime details of the underlying microscopic theory become unimportant: there exists a striking universality in thermodynamic behaviour. This universality now appears as possibly connected to the universality of gravitational attraction: all systems, whatever their non-gravitational interactions, display the same gravitational behaviour. If gravitation is indeed the manifestation of thermodynamic behaviour of a system that at the microscopic level is gravitation-free, then the universality of gravitation has the prospect of being explained in the same way as universality in thermodynamics.

The correspondence \textbf{b} in our scheme results from the combination of a surface-bulk correspondence at the micro-level and the thermodynamic limit (see Figure \ref{Micro-macro}). One interesting feature arising from this combination of ideas from holography and thermodynamics is, as we already remarked, that the correspondence \textbf{a} on the microscopic level could link two \emph{non}-gravitational theories, one in two dimensions and one in three. On the other hand, the correspondence \textbf{b} is between a non-gravitational thermodynamic surface theory and a theory of gravity in the bulk, as usual in holography (cf.\ the introduction of holography by 't Hooft, reported in section 2). In Verlinde's scheme the standard holographic relation can therefore be interpreted as arising from a more elementary non-gravitational holographic mapping \textbf{a}, combined with taking a thermodynamic limit.
The presentation in Figure \ref{Micro-macro} differs from the AdS/CFT case discussed earlier in that the microscopic theory of the former does not need to contain gravity, whereas usually in interpretations of AdS/CFT it is assumed that \emph{a2} is a microscopic theory of gravity. One naturally wonders whether an interpretation of AdS/CFT is also possible along the lines of what we just discussed. We will comment on this possibility in the next section and our conclusion.

Actually, an appeal to a microscopic theory on the bulk side (dealing with \emph{a2}) may not be necessary: such a theory plays no role in the argument for the emergence of gravity. The holographic reinterpretation of the thermodynamics on the screen suffices for the introduction of gravity, so it may be sufficient to look at the bulk counterpart of the surface theory in which the thermodynamic limit has already been taken. This raises the question of whether we have to assume a mapping between microscopic theories, \textbf{a} in the above, at all. In other words, we should consider the possibility that there might be a holographic mapping between bulk and surface theories only \emph{after} the thermodynamic limit has been taken, on the macro-level. In the next section we will discuss this possibility of a new interpretation in more detail. This discussion will lead us to an unusual view on holography, in which holography \emph{itself} emerges in the thermodynamic limit. This may also be relevant for AdS/CFT case.

\subsection{Emergence, holography, and thermodynamics}

As noted, two elements play an essential role in the correspondence between the surface theory and the bulk in Verlinde's scheme represented in Figure \ref{Micro-macro}: the holographic correspondence and the transition to the thermodynamic regime.
First we will add some comments on how holography and coarse-graining work together here and then we will discuss to what extent gravity and space can be said to \emph{emerge}.

As we have seen in section 3.2, in the case of an exact duality between surface and bulk we can distinguish different situations. There might be reasons to think that the two theories linked by the duality are structurally similar but still different: indeed, one is about a two-dimensional surface and the other about three-dimensional space. However, in the case in which an external point of reference is lacking and in which we cannot tell \emph{a priori} which quantities should be called spatial and how many spatial dimensions there fundamentally are, the meaning of physical quantities can only be given by the role they play in the theoretical framework. According to this `internal' viewpoint, which seems the appropriate one in the present context because the meaning of the surface theory is exactly the point under discussion, the fact that all observables, their (expectation) values and their mutual relations in the two respective theories stand in a bijective correspondence to each other means that we are dealing with two different formulations of one and the same physical theory. As argued in section 3.2, in both cases of exact duality, with and without an external viewpoint, the concept of emergence is not applicable. It follows that in the scenario of Figure \ref{Micro-macro}, with an assumed exact holographic duality, gravity and space cannot be said to emerge from holography.

On the other hand, that taking a thermodynamic limit can lead to emergence is a standard observation: The transition to thermodynamics opens up a new level of description that is characterized by new concepts and by patterns of physical behaviour that to a large extent are independent of the microscopic details of the underlying theory. In this sense gravity \emph{can} be said to emerge in Verlinde's scheme: as we have seen, it appears as an entropic force that has no counterpart on the micro-level. Its characteristics are independent of the details of the microscopic interactions and depend only on universal thermodynamic relations; thus, the universality of gravitation appears as a sign of its emergent thermodynamic character.

Verlinde writes that according to his proposal ``space is emergent through a holographic scenario.''\footnote{From the abstract in (Verlinde,  2011).} However, as noted, holography cannot be responsible for this emergence; if space emerges, it must be the thermodynamic limit that does the work. But is space really emergent in the scenario summarized in Figure \ref{Micro-macro}? It is true that the spatial coordinate in the bulk theory corresponds to a coarse-graining variable in the surface theory (or the number of renormalization steps that are taken), but this variable is just a \emph{parameter} that keeps track of the level of coarse-graining (see section 4.1). In other words, it is not itself a thermodynamic quantity. The parameter $x$ on the surface side is \emph{reinterpreted} as a spatial coordinate on the bulk side via the holographic connection, but as we have discussed in sect.\ 3.3.2, such a reinterpretation by itself does not lead to emergence. It therefore follows that although gravity can be said to emerge as a thermodynamic phenomenon, space itself does \emph{not} emerge in this scenario.

\begin{figure}[top]
\begin{center}
\includegraphics[height=6cm]{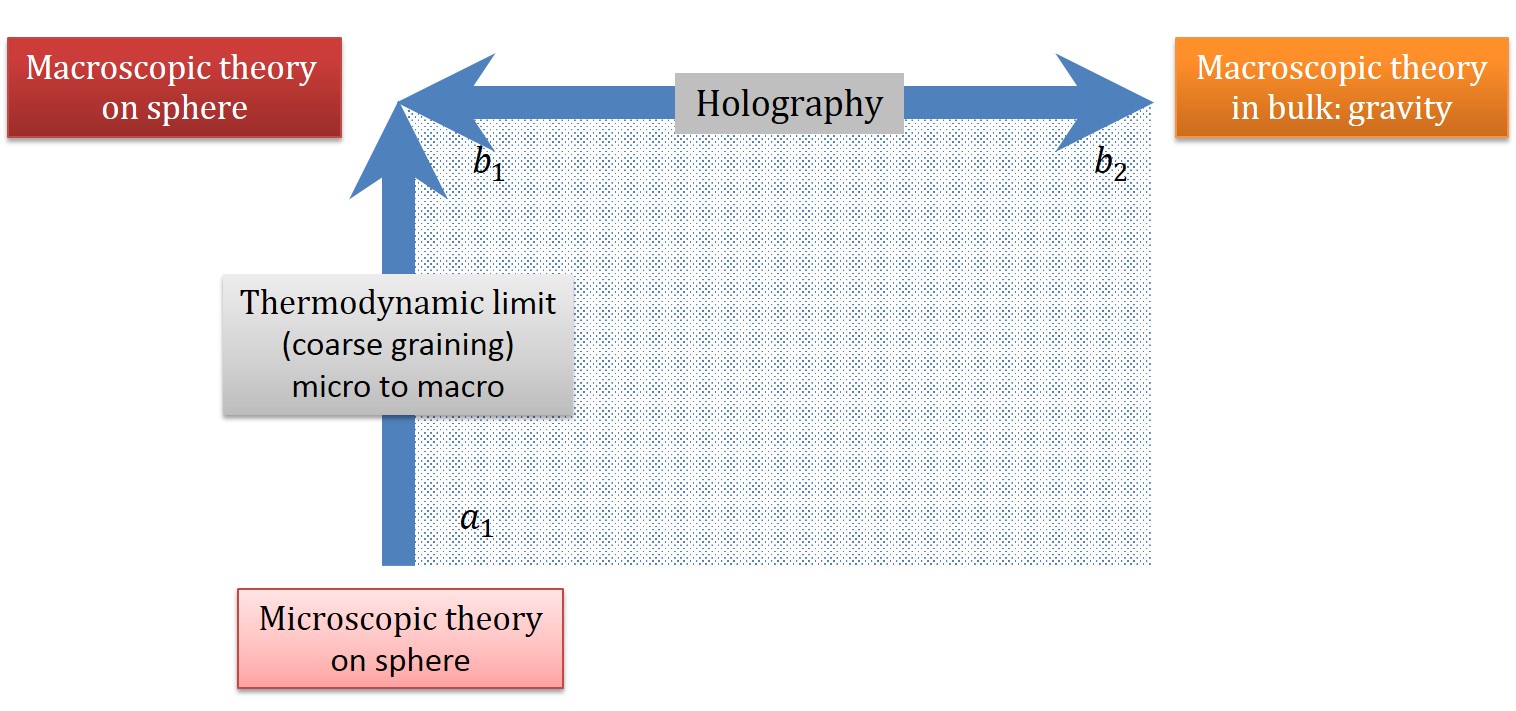}
\caption{Second interpretation of Verlinde's emergence scenario: thermodynamic emergence of space, gravity and holography without a microscopic bulk theory. }
\label{Verlinde-emergence}
\end{center}
\end{figure}

But as we pointed out at the end of section 4.2, invoking \emph{a2} is not indispensable for arriving at the gravitational system \emph{b2}. One may therefore consider an alternative reading of the scheme in which there is no holographic relation \textbf{a} and in which holography appears as a relation that only makes sense on the thermodynamic level: see Figure \ref{Verlinde-emergence}. The holographic relation {\bf b} is now not analyzable as the combined result of a microscopic holographic relation plus a thermodynamic limit: according to this new suggestion there simply \emph{is} no holography at the microscopic level---in a sense, holography becomes a thermodynamic phenomenon itself.

So the suggestion of Figure \ref{Verlinde-emergence} is to deny the existence of a bulk system \emph{a2}. This suggestion agrees with one of the motivating ideas behind Verlinde's approach, namely that there may be no need for a quantum theory of gravity at the microscopic level and for grand unification.\footnote{``The quest for unification of gravity with these other [quantum] forces of Nature, at a microscopic level, may [...] not be the right approach'', Verlinde writes, with reference to the many problems that this quantization approach has produced, (Verlinde, 2011), pp.\ 1-2.} This accords with the idea that there is no microscopic bulk theory \emph{of gravity}; but it would also agree with the idea that there is no microscopic three-dimensional bulk at all.

When we accept the analysis of the situation as depicted in Figure \ref{Verlinde-emergence}, the possibilities with respect to emergence change drastically. The holographic correspondence now only arises \emph{after}, and \emph{because}, we have taken the thermodynamic limit. In this case there is only a correspondence between the surface theory and a three-dimensional counterpart on the level of an effective thermodynamic description, and the existence of three-dimensional space need not be admitted on the micro-level. As a consequence, in this alternative scenario the thesis that ``space emerges together with gravity''\footnote{(Verlinde, 2011), on p. 2.} \emph{could} be justified.

Yet another representation of the situation now suggests itself. If we take the equality signs in relations (\ref{entropy}--\ref{Newton}) seriously, implying that there is an exact duality on the macroscopic level, then we should identify \emph{b1} and \emph{b2} (indeed, physicists refer to the correspondence here as given by a `dictionary', i.e.,\  by relations of synonymy). In this case, all that remains in our diagram is a `diagonal' arrow, connecting the microscopic \emph{a1} to the macroscopic \emph{b2} (Fig.\ 4). This arrow relates a lower dimensional quantum theory to a higher dimensional gravitational theory, so would stand for a `holographic' relation. Yet, this holographic relation would now \emph{include the limit from micro to macro}.
If we introduce this novel sense of holography, then it would have to be an example of an \emph{inexact} duality, as the thermodynamic component washes away the details of the quantum mechanical microstructure. This accords with our discussion of AdS/CFT, where we pointed out that only inexact dualities open up the possibility for emergence. This move also reminds us of the suggestion made in subsection 3.3.3 that AdS/CFT might exhibit emergence of gravity if coarse graining is included in the account. Finally, we suggested earlier that Verlinde does not properly distinguish holographic and thermodynamic aspects of emergence; however, on the reading just given, his discussion of emergence could be justified because holography and thermodynamics are now combined into one `emergence relation' (the single arrow in Fig.\ 4). The appearance of holography is then inseparably bound up with going from micro to macro descriptions.

\begin{figure}[top]
\begin{center}
\includegraphics[height=6cm]{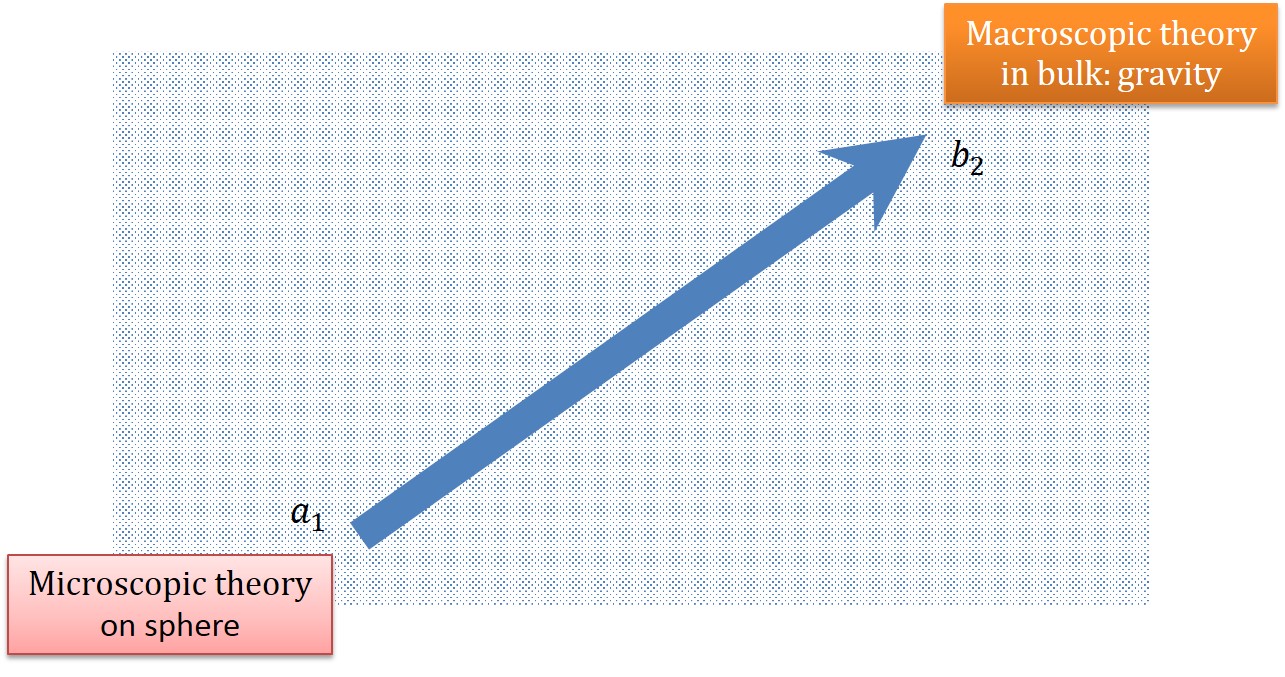}
\caption{Third interpretation of Verlinde's emergence scenario: thermodynamic emergence of space and gravity from a single  microscopic surface theory. The sinlge arrow includes the limit from micro to macro descriptions and `holographically' relates a  gravity theory in the bulk to a lower dimensional quantum theory.}
\label{Verlinde-emergence2}
\end{center}
\end{figure}

\section{Conclusion: emergence and holography}

We have reviewed three cases: 't Hooft's original holographic proposal, AdS/CFT, and Verlinde's
recent scheme. In 't Hooft's 1993 introduction of the holographic hypothesis there is no clear case for emergence, even though 't Hooft's text in places suggests a more fundamental status for the physics on the boundary. The original introduction of holography was programmatic and rather ambiguous in its interpretational aspects.

The case of AdS/CFT is more clear-cut, because in it the notion of holography is made more precise as a duality relation.
We have argued that if this duality is exact, as is generally expected and is suggested by calculations, there is no reason to consider one
of the two holographically related quantum theories as emergent from the other: bulk and boundary theories are two representations of one and the same theory.
Although many have expressed the intuition that the spacetime bulk is somehow emergent from the boundary field theory, this does not seem tenable when exact duality is accepted together with what we in section \ref{internalist} called an `internalist' viewpoint on theoretical terms. The only possible place for emergence in this case appears
to be the emergence of Einsteinian gravity, via RG flow, from an underlying microscopic theory that is not explicitly gravitational. This is, however, not the same thing as emergence of gravity from the boundary theory, but rather emergence due to coarse graining.

Continuing this line of thought, it is important to note that in studies of AdS/CFT the gravity side has
mostly been formulated in a semi-classical regime of small string length compared to the AdS radius, which means that a macroscopic limit has implicitly been taken.
It could be that only in this limit a spacetime point of view becomes applicable. In this case the holographic relation would arise
\emph{together} with the interpretation of certain degrees of freedom as gravitational. Gravity could thus be a manifestation of
thermodynamic behaviour and a microscopic quantum gravity theory would not be needed.

Seen from this angle, AdS/CFT might fit the same scheme as we have proposed in connection with Verlinde's approach in Figure \ref{Verlinde-emergence} and Figure 4. In the latter case, understanding gravity as emergent is fairly unproblematic: gravity has a thermodynamic origin, and the gravitational force is an entropic force without a corresponding microphysical interaction. One might say that holography, spacetime and gravity here emerge together in the thermodynamic limit of a microphysical theory without gravity on a screen. Could one adopt this kind of interpretation also for AdS/CFT? Verlinde seems to say as much when he writes that the gravitational side of AdS/CFT should not be seen as ``independently defined''; he compares referring to gravity in AdS/CFT to ``using stress tensors in a continuous medium half a century before knowing about atoms."\footnote{(Verlinde, 2011), p.\ 21. Other authors, in different contexts than string theory's AdS/CFT scenario, have expressed similar views about the possibility of an intrinsically coarse grained nature of gravity; see e.g. the references in note \ref{alternatives}.}
This appears to go in the direction of a critical reinterpretation of AdS/CFT in which one abandons the notion of microscopic gravity.
Of course, Verlinde's own scheme takes an explicit thermodynamical step, which is not present in standard discussions of AdS/CFT. However, as we have seen in section \ref{renorm1}, moving inward in the bulk is in AdS/CFT related to coarse-graining renormalization transformations on the boundary which suggests a relation with statistical physics and thermodynamics here as well (this aspect of AdS/CFT in fact has been one of the inspirations for Verlinde's scenario).\footnote{Verlinde (2011, pp.\ 20-25) sees a number of other reasons that support the idea of abandoning the notion of microscopic gravity theories. He points to UV/IR relations (of which AdS/CFT is only one example, another being open/closed string duality) as indications that long range gravitational forces seem to know about high energy, short distance physics. This could be a sign that gravity should not be considered as an independently defined quantum force on the micro-level. Another indication could be seen in the set of relations known collectively as `black hole thermodynamics', which originated in the 1970s from work by Jakob Bekenstein, Stephen Hawking and others; see e.g.\ (Bekenstein 1973), (Hawking et al., 1973). Here we find relations between black hole quantities, originally defined exclusively in terms of Einsteinian gravity, that completely mimic the familiar thermodynamic laws. A final reason for the hypothesis that there is no microscopic gravity is the universality of gravity mentioned before.}

We think that the preceding analysis explains how gravitation and space may be understood as emergent in holographic scenarios. If gravity is emergent, this has far-reaching ramifications: for example, if gravity is an entropic force, then there is no point in
looking for a microscopic quantum theory  of gravity, or in seeking gravity's unification with other microscopic forces.
Furthermore, if gravity is a thermal phenomenon, one may expect fluctuations around the macroscopic equilibrium state; that is, small deviations from the
Einstein theory.\footnote{Such deviations could also arise in a theory that successfully quantizes gravity, so they are not specific to the Verlinde scheme.}
Our analysis shows that it is the transition to the thermodynamic regime, and not holography, that produces the emergent properties in these new scenarios, just as in traditional and  familiar examples of emergence in physics. A novel conceptual possibility that has arisen in this investigation is that in AdS/CFT-like accounts the holographic relation itself may emerge along with gravity.

\section*{Acknowledgments}

We are grateful to Jeremy Butterfield, Diego Hofman, and two anonymous referees, for their commentary on this article. We also thank Erik Verlinde for feedback on our paper and fruitful discussions; and audiences in Munich, Florence, Chicago and Potsdam for their comments.

\section*{References}\addcontentsline{toc}{section}{References}

Acu\~{n}a, P. and D. Dieks, (2014). ``Another look at empirical equivalence and underdetermination of theory choice", \emph{European Journal for Philosophy of Science}, 4, 153-180.\\
\\
Aharony, O.,  S.S. Gubser, J.M. Maldacena, H. Ooguri, and Y. Oz. (2000). ``Large \emph{N} field theories, string theory and gravity'',  \emph{Physics Reports}, 323(3-4), 183-386.  [hep-th/9905111].\\
  %%CITATION = HEP-TH/9905111;%%
\\
Bardeen, J.M., B. Carter, and S.W. Hawking. (1973). ``The four laws of black hole mechanics", \emph{Communications in Mathematical Physics}, 31(2), 161-170.\\
\\
 Barcelo, C., Visser, M., Liberati, S. (2001). ``Einstein gravity as an emergent phenomenon?'',
  {\it International Journal of Modern Physics D} 10(6), 799-806.
  [gr-qc/0106002].\\
  %%CITATION = GR-QC/0106002;%%
\\
Batterman, R.W. (2011). ``Emergence, singularities and symmetry breaking'', \emph{Foundations of Physics}, 41(6), 1031-1050.\\
\\
Bekenstein, J. (1973). ``Black holes and entropy", \emph{Physical Review D}, 7(8), 2333-2346.\\
\\
Belot, G., J.~Earman, and L.~Ruetsche (1999). ``The Hawking information loss paradox: the anatomy of controversy'', \emph{The British Journal for the Philosophy of Science}, 50(2), 189-229.\\
\\
Berenstein, D., and R. Cotta. (2006). ``Aspects of emergent geometry in the AdS/CFT context'', \emph{Physical Review D}, 74(2), 026006. [hep-th/0605220].\\
\\
Bianchi, M. (2001). ``(Non-)perturbative tests of the AdS/CFT correspondence'', \emph{Nuclear Physics B. Proceedings Supplements}, 102-103, 56-64. [hep-th/0103112].\\
  %%CITATION = HEP-TH/0103112;%%
\\
Boer, J. de, E.P. Verlinde, and H.L. Verlinde, (2000). ``On the holographic renormalization group'', \emph{Journal of High Energy Physics}, 08 003.
  [hep-th/9912012].\\
  %%CITATION = HEP-TH/9912012;%%
\\
Bouatta, N. and Butterfield, J. (2015), ``On emergence in gauge theories at the 't Hooft limit'', \emph{European Journal for Philosophy of Science}, 5(1), 55-87. [arXiv:1208.4986 [physics.hist-ph]].\\
  %%CITATION = ARXIV:1208.4986;%%
\\
Bousso, R. (2002). ``The holographic principle'', \emph{Reviews of Modern Physics}, 74(3), 825-874.  [hep-th/0203101].\\
  %%CITATION = HEP-TH/0203101;%%
\\
Butterfield, J. (2011a). ``Emergence, reduction and supervenience: a varied landscape'', \emph{Foundations of Physics}, 41(6), 920-959. \\
\\
Butterfield, J. (2011b). ``Less is different: emergence and reduction reconciled'', \emph{Foundations of Physics}, 41(6), 1065-1135.\\
\\
Castellani, E. (2010). ``Dualities and intertheoretic relations", pp. 9-19 in: Suarez, M., M. Dorato and M. Red\'{e}i (eds.). \emph{EPSA Philosophical Issues in the Sciences}. Dordrecht: Springer.\\
\\
Csaki, C., H. Ooguri, Y. Oz, and J. Terning. (1999). ``Glueball mass spectrum from supergravity'', \emph{Journal of High Energy Physics}, 01 017.
  [hep-th/9806021].\\
  %%CITATION = HEP-TH/9806021;%%
\\
Cubrovi{\'c}, M., J. Zaanen, K. Schalm. (2009). ``String theory, quantum phase transitions, and the emergent fermi liquid'', \emph{Science}, 325(5939), 439-444. [arXiv:0904.1993 [hep-th]].\\
\\
Dom\`enech, O., M. Montull, A. Pomarol, A. Salvio, and P.J. Silva. (2010). ``Emergent gauge fields in holographic superconductors'', \emph{Journal of High Energy Physics}, 2010 033.  [arXiv:1005.1776 [hep-th]].\\
\\
Dongen, J. van. (2010). \emph{Einstein's unification}. Cambridge: Cambridge University Press. \\
\\
Dongen, J. van, and S.~de Haro. (2004). ``On black hole complementarity'', \emph{Studies in History and Philosophy of Science Part B: Studies in History and Philosophy of Modern Physics}, 35(3), 509-525.\\
\\
Drukker, N.,  Marino, M., and  Putrov, P. (2011). ``Nonperturbative aspects of ABJM theory'',  \emph{Journal of High Energy Physics},  2011 141. [arXiv:1103.4844 [hep-th]].\\
  %%CITATION = ARXIV:1103.4844;%%
\\
Fisher, M.E. (1998). ``Renormalization group theory: Its basis and formulation in statistical physics'', \emph{Reviews of Modern Physics}, 70(2), 653-681.\\
\\
Gaite, J.C. and O'Connor, D. (1996). ``Field theory entropy, the \emph{H} theorem, and the renormalization group'', \emph{ Physical Review D}, 54, 5163-5173.
  [hep-th/9511090].\\
  %%CITATION = HEP-TH/9511090;%%
\\
Giddings, S.B. (2011). ``Is string theory a theory of quantum gravity?'', \emph{Foundations of Physics}, 43(1), 115-139.  [arXiv:1105.6359 [hep-th]].\\
  %%CITATION = ARXIV:1105.6359;%%
\\
Green, M.B. (1999). ``Interconnections between type II superstrings, M theory and \emph{N}=4 supersymmetric Yang-Mills'', pp. 22-96 in: Ceresole, A., C. Kounnas, D. L\"{u}st, and S. Theisen (eds.). \emph{Quantum aspects of gauge theories, supersymmetry and unification}. Berlin, Heidelberg: Springer.  [hep-th/9903124].\\
  %%CITATION = HEP-TH/9903124;%%\\
\\
Green, M.B., Schwarz,  J.H. and  Witten, E. (1987). \emph{Superstring theory. Vol. 1: Introduction}. Cambridge: Cambridge University Press.\\
\\
Gubser, S.S., Klebanov,  I.R.  and  Polyakov, A.M.. (1998). ``Gauge theory correlators from noncritical string theory'',  \emph{Physics Letters B}, 428(1-2), 105-114.  [hep-th/9802109].\\
  %%CITATION = HEP-TH/9802109;%%
\\
Haro, S. de, Skenderis, K. and Solodukhin, S. (2001). ``Holographic reconstruction of spacetime and renormalization in the AdS/CFT correspondence", \emph{Communications in Mathematical Physics}, 217(3), 595-622. [hep-th/0002230].\\
\\
Hartmann, S. (2001). ``Effective field theories, reductionism and scientific explanation'',  \emph{Studies in History and Philosophy of Science Part B: Studies in History and Philosophy of Modern Physics}, 32(2), 267-304.\\
\\
Hartnoll, S.A.,Herzog,  C.P., and Horowitz, G.T.  (2008). ``Building a holographic superconductor'', \emph{Physical Review Letters}, 101(3), 031601.\\
\\
Hawking, S.W. (1976). ``Breakdown of predictability in gravitational collapse'', \emph{Physical Review D}, 14(10), 2460-2473.\\
\\
Hertog, T. and Horowitz, G. T.  (2005). ``Holographic description of AdS cosmologies'',
\emph{Journal of High Energy Physics}, 0504 005.
  [hep-th/0503071].
  %%CITATION = HEP-TH/0503071;%%
\\
\\
Hooft, G. 't. (1985). ``On the quantum structure of a black hole'', \emph{Nuclear Physics B}, 256, 727-745.\\
  %%CITATION = NUPHA,B256,727;%%
\\
Hooft, G. 't. (1993). ``Dimensional reduction in quantum gravity'', in: Ali, A., J. Ellis and S. Randjbar-Daemi, \emph{Salamfestschrift}. Singapore: World Scientific. [gr-qc/9310026].\\
  %%CITATION = GR-QC/9310026;%%\\
\\
Hooft, G. 't. (1996). ``The scattering matrix approach for the quantum black hole: an overview'', \emph{International Journal of Modern Physics A}, 11(26), 4623-4688. [gr-qc/9607022].\\
  %%CITATION = GR-QC/9607022;%%
\\
Hooft, G. 't. (1999). ``Quantum gravity as a dissipative deterministic system'', \emph{Classical and Quantum Gravity}, 16(10), 3263-3279.  [gr-qc/9903084].\\
  %%CITATION = GR-QC/9903084;%%\\
\\
Horowitz, G.T. (2005). ``Spacetime in string theory'',  \emph{New Journal of Physics},  7 201.  [gr-qc/0410049].  \\
%%CITATION = GR-QC/0410049;%%
\\
Hu, B.~L.~(2009). ``Emergent/quantum gravity: Macro/micro structures of spacetime'',
{\it  Journal of Physics Conference Series} 174(1), 012015
  [arXiv:0903.0878 [gr-qc]].\\
  %%CITATION = ARXIV:0903.0878;%%
\\
Jacobson, T. (1995). ``Thermodynamics of spacetime: The Einstein equation of state", \emph{Physical Review Letters}, 87(7), 1260-1263.\\
\\
 Konopka, T., Markopoulou, F., Severini, S. (2008). ``Quantum Graphity: A Model of emergent locality,''
{\it  Physical Review D} 77, 104029
  [arXiv:0801.0861 [hep-th]].\\
  %%CITATION = ARXIV:0801.0861;%%
\\
Laudan, L. and L. Leplin. (1991). ``Empirical equivalence and underdetermination", \emph{The Journal of Philosophy}, 88(9), 449-472.\\
\\
Maldacena, J. (1998). ``The large \emph{N} limit of superconformal field theories and supergravity'',  \emph{Advances in Theoretical and Mathematical Physics} 2, 231-252.
  [hep-th/9711200].\\
  %%CITATION = HEP-TH/9711200;%%
\\
Matsubara, K. (2013).  ``Realism, underdetermination and string theory dualities'', \emph{Synthese}, 190, 471-489.\\
\\
McGreevy, J. (2010). ``Holographic duality with a view toward many-body physics", \emph{Advances in High Energy Physics}, 723105. [arXiv:0909.0518].\\
\\
Oriti, D.~(2014). ``Disappearance and emergence of space and time in quantum gravity'',
 {\it  Studies in History and Philosophy of Science Part B: Studies in History and Philosophy of Modern Physics}, 46,  186-199.
  [arXiv:1302.2849 [physics.hist-ph]].\\
  %%CITATION = ARXIV:1302.2849;%%
\\
Padmanabhan, T.~(2015). ``Emergent gravity paradigm: Recent progress'',
 {\it  Modern Physics Letters A} 30(03n04),  1540007.
  [arXiv:1410.6285 [gr-qc]].\\
  %%CITATION = ARXIV:1410.6285;%%
\\
Rickles, D. (2011). ``A philosopher looks at string dualities'',  \emph{Studies in History and Philosophy of Science Part B: Studies in History and Philosophy of Modern Physics}, 42(1), 54-67.\\
\\
Rickles, D. (2012). ``AdS/CFT duality and the emergence of spacetime'', \emph{Studies in History and Philosophy of Science Part B: Studies in History and Philosophy of Modern Physics}, 44(3), 312-320.\\
\\
Seiberg, N. (2007). ``Emergent spacetime'', pp. 163-213 in: Gross, D., M. Henneaux, and A. Sevrin (eds.). \emph{The quantum structure of space and time. Proceedings of the 23rd Solvay conference on physics.} Singapore: World Scientific. [hep-th/0601234].  \\
%%CITATION = HEP-TH/0601234;%%
\\
Smolin, L. (2007). \emph{The Trouble With Physics: The Rise of String Theory, The Fall of a Science, and What Comes Next}. New York: Mariner Books.\\
\\
Stachel, J. (1993). ``The other Einstein: Einstein contra field theory", \emph{Science in Context}, 6(1), 275-290.\\
\\
Susskind, L., L. Thorlacius, and J.~Uglum. (1993). ``The stretched horizon and black hole complementarity'', \emph{Physical Review D}, 48(8),  3743-3761.  [hep-th/9306069].\\
  %%CITATION = HEP-TH/9306069;%%
\\
Susskind, L. (1995). ``The world as a hologram'',  \emph{Journal of Mathematical Physics},  36(11), 6377-6396.  [hep-th/9409089].\\
  %%CITATION = HEP-TH/9409089;%%
\\
Susskind, L. and E. Witten. (1998). ``The holographic bound in anti-de Sitter space'', 10 pp. [hep-th/9805114].\\
  %%CITATION = HEP-TH/9805114;%%
\\
Swingle, B. (2012), ``Entanglement renormalization and holography'',
  \emph{Physical  Review D}, 86(6), 065007.
  [arXiv:0905.1317 [cond-mat.str-el]].\\
  %%CITATION = ARXIV:0905.1317;%%
\\
Teh, N.J. (2013). ``Holography and emergence'', \emph{Studies in History and Philosophy of Science Part B: Studies in History and Philosophy of Modern Physics}, 44(3), 300-311.\\
\\
Verlinde, E. (2011). ``On the origin of gravity and the laws of Newton'', \emph{Journal of High Energy Physics}, 2011 029.  [arXiv:1001.0785 [hep-th]].\\
  %%CITATION = ARXIV:1001.0785;%%
\\
Witten, E. (1998a). ``Anti-de Sitter space and holography'', \emph{Advances in Theoretical and Mathematical Physics}, 2, 253-291. [hep-th/9802150].\\
  %%CITATION = HEP-TH/9802150;%%
\\
Witten, E. (1998b). ``Anti-de Sitter space, thermal phase transition, and confinement in gauge theories,''   \emph{Advances in Theoretical and Mathematical Physics}, 2, 505-532.  [hep-th/9803131].\\
  %%CITATION = HEP-TH/9803131;%%
\\
Zwiebach, B. (2004). \emph{A first course in string theory}. Cambridge: Cambridge University Press.

\end{document}